\documentclass[aps,pre,showpacs,twocolumn,showkeys,amssymb,amsmath,superscriptaddress,reprint]{revtex4-1}

\usepackage{amsmath}
\usepackage{graphicx}
\usepackage{color}

\usepackage{verbatim}

\usepackage{amssymb}
\usepackage{bm}

\DeclareMathOperator{\Tr}{Tr}

\begin{document}

\title{The metastable minima of the  Heisenberg spin glass in a random magnetic field}

\author{Auditya Sharma}
\affiliation{Department of Physics, Indian Institute of Science Education and Research, Bhopal, India}
\author{Joonhyun Yeo}
\affiliation{Department of Physics,
Konkuk University, Seoul 143-701, Korea}

\author{M.~A.~Moore}
\affiliation{School of Physics and Astronomy, University of Manchester,
Manchester M13 9PL, UK}


\begin{abstract}
We  have  studied  zero  temperature metastable  minima  in  classical
$m$-vector  component spin  glasses in  the presence  of $m$-component
random  fields for  two  models,  the  Sherrington
Kirkpatrick (SK)  model and the  Viana Bray (VB)  model. For the  SK model we have calculated  analytically its complexity
(the log of the number of minima) for both the annealed case where one
averages the number  of minima before taking the log  and the quenched
case where one  averages the complexity itself, both  for fields above
and below the  de Almeida Thouless (AT) field, which  is finite for $m
>2$. We  have done numerical  quenches starting from a  random initial
state (infinite temperature state) by  putting spins parallel to their
local fields  until there  is no  further decrease  of the  energy and
found that  in zero  field it  always produces  minima which  have zero
overlap with each other. For the $m=2$ and $m=3$ cases in the SK model
the final  energy reached in  the quench is  very close to  the energy
$E_c$  at  which the  overlap  of  the  states would  acquire  replica
symmetry breaking  features. These minima have  marginal stability and
will have  long-range correlations  between them. In  the SK  limit we
have analytically studied the density of states $\rho(\lambda)$ of the
Hessian matrix in  the annealed approximation.  Despite  the fact that
in the presence of a random  field there are no continuous symmetries,
the spectrum extends down to zero with the usual $\sqrt{\lambda}$ form
for the density of states for fields below the AT field. However, when
the random field  is larger than the  AT field, there is a  gap in the
spectrum which  closes up as  the AT field  is approached. The VB model 
behaves differently and seems rather similar to studies of the three dimensional Heisenberg spin glass in a random vector field.
\end{abstract}

\maketitle

\section{Introduction}
\label{sec:introduction} 
In  recent years  there  has  been a  resurgence  of  interest in  the
properties  of metastable  states, due  mostly to  the studies  of the
jammed   states   of   hard    sphere   systems;   see for   reviews  
Refs. \onlinecite{charbonneau16,  baule16}. There  are many  topics to
study,  including  for example  the  spectrum  of small  perturbations
around  the metastable  state,  i.e. the  phonon  excitations and  the
existence of  a boson peak,  and whether the Edwards  hypothesis works
for these states. In this paper we shall study some of these topics in
the context of classical Heisenberg  spin glasses both in the presence 
and absence of a random magnetic
field. Here the metastable states which we study  are just the minima of the Hamiltonian, and so are well-defined outside the mean-field limit. It  has  been  known  for some  time  that  there  are  strong
connections    between   spin    glasses   and    structural   glasses
~\cite{tarzia2007glass,fullerton2013growing,  moore06}.  It  has  been
argued in very recent work~\cite{baity2015soft}  that the study of the
excitations  in classical Heisenberg  spin glasses  provides the  opportunity to
contrast     with      similar     phenomenology      in     amorphous
solids~\cite{wyart2005geometric,   charbonneau15}. The minima and excitations about the minima    in
Heisenberg spin glasses have been studied for many years \cite{bm1981,
  yeo04, bm1982} but only in the absence of external fields.

 In  Sec.  \ref{sec:models}  we define  the models  to be  studied as
 special  cases of  the  long-range one - dimensional $m$-component vector  spin
 glass  where the  exchange  interactions $J_{ij}$  decrease with  the
 distance   between   the   spins   at    sites   $i$   and   $j$   as
 $1/r_{ij}^{\sigma}$. The spin $\mathbf{S}_i$ is an $m$-component unit vector. $m=1$ corresponds to the Ising model, $m=2$ corresponds to the XY model and $m=3$ corresponds to the Heisenberg model. By  tuning the parameter $\sigma$,  one can have
 access to  the Sherrington-Kirkpatrick (SK)  model and on  dilution to
 the  Viana-Bray (VB)  model, and  indeed to  a range  of universality
 classes     from     mean-field-type      to     short-range     type
 \cite{leuzzi2008dilute},  although in  this  paper  only two  special
 cases are studied;  the SK model and the Viana-Bray  model. We intend
 to study the cases which correspond to short-range models in a future
 publication.

In Sec. \ref{sec:metastability} we have used numerical  methods to learn about
the metastable minima of the SK model and the Viana Bray model.  Our main procedure for finding the minima is
to start from  a random configuration of
spins  and then align each  spin  with the  local field
produced by its neighbors and the external random field, if present. The process is
continued until  all spins are  aligned with their local  fields. This
procedure finds local minima of  the Hamiltonian. In the thermodynamic
limit, the  energy per spin $\varepsilon$ of these states reaches  a characteristic
value, which is  the same for  almost all realization  of the bonds and random external fields, but slightly dependent  on the dynamical algorithm used  for selecting the spin  to be
flipped e.g. the ``polite''  or ``greedy''  or Glauber dynamics or the sequential algorithm used in the numerical work in this paper
\cite{newman:99,parisi:95}. 
   In the
context of  Ising spin glasses in zero random fields  such states  were first 
studied by  Parisi  \cite{parisi:95}.   For Ising spins these
dynamically  generated states  are an  unrepresentative subset  of the
totality  of the  one-spin  flip stable  metastable  states, which  in
general have  a distribution  of local fields  $p(h)$ with $p(0)$ is
finite  \cite{roberts:81},  whereas  those generated  dynamically  are
marginally stable and  have $p(h) \sim h$, just like  that in the true
ground state  \cite{yan:15}. Furthermore  these states have  a trivial
overlap with  each other: $P(q)= \delta(q)$ \cite{parisi:95};  there is
no  sign of  replica  symmetry breaking  amongst  them. Presumably  to
generate  states which  show  this  feature one  needs  to start  from
initial spin configurations drawn from  a realization of the system at
a temperature where broken replica  symmetry is already present before
the quench.

 Because the initial state is random, one would also expect for vector
 spin glasses that  the states reached after the  quench from infinite
 temperature  would  have  only  a trivial  overlap  with  each  other
 \cite{newman:99}  and  this  is  indeed  found  to  be  the  case  in
 Sec. \ref{sec:overlap}.  We have studied  the energy which is reached
 in the quench for both the $m=2$ and $m=3$ SK models but for the case
 of zero applied  random field and in  both cases it is  very close to
 the  energy $E_c$  which marks  the boundary  above which  the minima
 where spins are parallel to  their local fields have trivial overlaps
 with each  other, while below it  the minima have overlaps  with full
 broken   replica  symmetry   features   \cite{bm:81a,  bm1981}.    In
 Ref. \onlinecite{bm1981}  the number $N_S(\varepsilon)$ of  minima of
 energy $\varepsilon$ was calculated for  the case of zero random field
 in the SK model and in fact it  is only for this model and zero field
 that   the  value   of  $E_c$   is   available.   That   is  why   in
 Sec. \ref{sec:marginal}  only this case was  studied numerically. The
 work in Sec. \ref{sec:SKanalytic} was the start of an attempt to have
 the same information in the presence of random vector fields.

 The number of minima $N_S(\varepsilon)$  is exponentially large so it
 is useful  to study  the complexity  defined as  $g(\varepsilon)= \ln
 N_S(\varepsilon)/N$,  where  $N$  is  the  number  of  spins  in  the
 system. Despite  the fact  that minima  exist over  a large  range of
 values of $\varepsilon$  a quench by a particular  algorithm seems to
 reach  just  the   minima  which  have  a   characteristic  value  of
 $\varepsilon$. What is striking is that this characteristic value is close
 to  the energy  $E_c$ at  which  the minima  would no  longer have  a
 trivial overlap  with each other  but would start to  acquire replica
 symmetry  breaking features,  at least  for  the $m=2$  and $m=3$  SK
 models in zero field.  The states reached in the quenches are usually
 described   as   being   marginally  stable   \cite{muller:15}.   The
 coincidence of the energy obtained in the numerical quenches with the
 analytically calculated  $E_c$ suggests that  long-range correlations
 normally associated with  a continuous transition will  also be found
 for  the quenched  minima  since  such features  are  present in  the
 analytical work at $E_c$ \cite{bm:81a}.   In the Ising case the field
 distribution $p(h)$  produced in  the quench  is very  different from
 that assumed when determining $E_c$, and the quenched state energy at
 $\approx -0.73$ was so far below from the Ising value of $E_c=-0.672$
 that the connection of its marginality to the onset of broken replica
 symmetry has been  overlooked. We believe that  the identification of
 the energy  $E_c$ reached  in the  quench with  the onset  of replica
 symmetry breaking in the overlaps of the minima is the most important
 of our results.

  In Sec.  \ref{sec:SKanalytic} we present  our analytical work on the
  $m$-component SK  model in the  presence of an  $m$-component random
  field.    It  has   been  shown   that  in   the  mean-field   limit
  ~\cite{sharma2010almeida} that  under the  application of  a random
  magnetic field,  of variance  $h_r^2$, there  is a  phase transition
  line in the $h_r - T$  plane, the so-called de Almeida-Thouless (AT)
  line,  across which  the  critical  exponents lie  in  the Ising  AT
  universality class.   Below this  line, the  ordered phase  has full
  replica symmetry  breaking.  This  ordered phase  is similar  to the
  Gardner  phase  expected  in high-dimensional  hard  sphere  systems
  \cite{charbonneau16}.  In  Sec.  \ref{sec:SKanalytic} we  study  the
  minima of  the Heisenberg  Hamiltonian in the  presence of  a random
  vector field.  In the presence  of such  a field the  Hamiltonian no
  longer has any  rotational invariances so one might  expect there to
  be big changes in the excitations about the minimum as there will be
  no Goldstone modes in the system.

  We start Sec.  \ref{sec:SKanalytic} by studying the  number of local
  minima $N_S(\varepsilon)$  of the Hamiltonian which  have energy per
  spin  of   $\varepsilon$.   The  calculation  within   the  annealed
  approximation, where one  calculates the field and  bond averages of
  $N_S(\varepsilon)$ is  just an extension of  the earlier calculation
  of Bray  and Moore  for zero random  field \cite{bm1981}.   When the
  random field $h_r> h_{AT}$, where $h_{AT}$ is the field at which the
  AT transition  occurs, the complexity is  zero, but $g(\varepsilon)$
  becomes non-zero  for $h_r <  h_{AT}$.  When  it is non-zero,  it is
  thought  better to  average the  complexity itself  over the  random
  fields and bonds  so that one recovers results likely  to apply to a
  typical  sample.  We  have   attempted  to  calculate  the  quenched
  complexity $g$ for  the SK model in the presence  of a random field.
  The presence  of this random  field greatly complicates  the algebra
  and  the calculations  in Sec.  \ref{sec:quenched} and  the Appendix
  really just  illustrate the  problems that  random fields  pose when
  determining the quenched average but do not overcome the algebraic
  difficulties.
 
  The annealed approximation  is much simpler 
 and using it we have calculated the density of states $\rho(\lambda)$
 of  the Hessian  matrix associated  with the  minimum for the SK model. 
   When  $h_r  >  h_{AT}$  there is  a  gap
 $\lambda_0$ in  the spectrum  below which  there are  no excitations.
 $\lambda_0$  tends to  zero  as $h_r  \to h_{AT}$.   For  $m \ge  4$,
 $\rho(\lambda)   \sim  \sqrt{\lambda-\lambda_0}$   as  $\lambda   \to
 \lambda_0$. For  $m =3$  the square root  singularity did  not occur,
 much to our  surprise. For $h_r < h_{AT}$,  the square-root singularity
 applies for  all $m >  2$ with  $\lambda_0=0$. Thus in  the low-field
 phase, despite  the fact that  in the  presence of the  random fields
 there  are  no continuous  symmetries  in  the  system and  hence  no
 Goldstone   modes,    there   are   massless   modes    present.   In
 Sec.  \ref{sec:density} we  present numerical  work which  shows that
 even for $h_r < h_{AT}$ when the annealed calculation of the density of
 states  of the  SK  model  cannot be  exact,  it  nevertheless is  in
 good agreement with our numerical data.

 We   have   also  calculated in Sec. \ref{sec:spinglasssusceptibility}  the   zero   temperature  spin   glass
 susceptibility $\chi_{SG}$  for $h_r >  h_{AT}$ for the SK  model and
 find that for all $m > 2$ it diverges to infinity as $h_r \to h_{AT}$
 just as is  found at finite  temperatures \cite{sharma2010almeida}.

For the SK  model, because the complexity is zero  for $h_r > h_{AT}$,
the  quench  produces states  sensitive  to  the  existence of  an  AT
field. The quench then goes to a state which is the ground state or at
least  one very  like  it. The  AT  field  is a  feature  of the  true
equilibrium state  of the system,  which in our  case is the  state of
lowest  energy.  In  Sec.  \ref{sec:spinglasssusceptibility}  we  have
studied  a  ``spin glass  susceptibility''  obtained  from the  minima
obtained in our numerical quenches and  only for the SK model is there
evidence for a diverging spin  glass susceptibility. For the VB model,
there is no  sign of any singularity in the  spin glass susceptibility
defined  as an  average over  the states  reached in  our quench  from
infinite temperature, but we cannot  make any statement concerning the
existence of an  AT singularity in the true ground  state. This is the
problem studied in Ref. \onlinecite{lupo:16}.

       Finally       in
Sec.  \ref{sec:conclusions}  we summarise our main results and make some suggestions for further research.

\section{Models}
\label{sec:models}
The Hamiltonians  studied in this paper are generically of the form
\begin{equation}
\mathcal{H} = -m\sum_{\langle i, j \rangle} J_{ij} \mathbf{S}_i \cdot \mathbf{S}_j - \sqrt{m}\sum_i \mathbf{h}_i \cdot \mathbf{S}_i \, ,
\label{Ham}
\end{equation}
where the $\mathbf{S}_i$, $i = 1, 2, \cdots, N$,
are classical $m$-component vector spins of unit length. This form of writing the Hamiltonian allows for easy comparison against a Hamiltonian where the spins are normalized to have length $\sqrt{m}$. We are particularly interested in Heisenberg spins, for which $m=3$. 
The magnetic fields $h_i^\mu$, where $\mu$ denotes a Cartesian spin component,
are chosen to be
independent Gaussian random fields, uncorrelated between
sites, with zero mean, which satisfy
\begin{equation}
[ h_i^\mu h_j^\nu]_{av} = h_r^2\, \delta_{ij}\, \delta_{\mu\nu} \, .
\label{hs}
\end{equation}
The notation $[\cdots]_{av}$ indicates an average over the quenched disorder and the magnetic fields.

We shall study two models, the Sherrington-Kirkpatrick (SK) model and the Viana-Bray  (VB) model. Both are essentially mean-field models. In the Sherrington-Kirkpatrick model, the bonds $J_{ij}$ couple all pairs of sites and are drawn from a Gaussian distribution with zero mean and the variance $1/(N-1)$. 

The Viana-Bray model can be regarded as a special case of a diluted one-dimensional model where the sites are arranged around a ring.
The procedure to determine the bonds $J_{ij}$ to get the diluted model is as specified in Refs.  \onlinecite{leuzzi2008dilute,sharma2011phase,sharma2011almeida}.
The probability of there being a non-zero interaction between sites $(i,j)$ on the ring falls off with distance as a power-law, and when an
interaction does occur, its variance is independent of $r_{ij}$. The mean number of non-zero bonds from a site is fixed to be $z$. 
To generate the set of pairs $(i,j)$ that have an interaction with the desired probability
the spin $i$ is chosen randomly, and then $j \ (\ne i)$ is chosen at distance $r_{ij}$ with probability 
\begin{equation}
p_{ij} = \frac{r_{ij}^{-2\sigma}}{\sum_{j\, (j\neq i)}r_{ij}^{-2\sigma}} \, ,
\end{equation}
where $r_{ij}=\frac{N}{\pi}\sin\left[\frac{\pi}{N}(i-j)\right]$ is the
length of the chord between the sites $i,j$ when all the sites are
put on a circle.  If $i$ and $j$ are already connected, the process is
repeated until a pair which has not been connected before is
found. The sites $i$ and $j$ are then connected with an interaction
picked from a Gaussian interaction whose mean is zero and whose
standard deviation is set to $J \equiv 1$.  This process is repeated
precisely $N_b = z N / 2 $ times. This procedure automatically gives
$J_{ii} = 0$. Our work concentrates on the case where the coordination number is fixed at $z=6$ to mimic the $3$-d cubic scenario.
The SK limit ($z=N-1, \sigma = 0$) is a special case of this model, as is the VB model which also has  $\sigma = 0$,  but
the coordination number $z$  has (in this paper)  the value  $6$. The advantage of the one-dimensional long-range  model for numerical studies is that by simply tuning the value of $\sigma$ one can mimic the properties of finite dimensional systems~\cite{leuzzi2008dilute,sharma2011phase,sharma2011almeida} and we have already done some work using this device. However, in this paper we only report on our work on the SK and VB models.

\section{Numerical studies of the minima obtained by quenching}
\label{sec:metastability} 

In this section we present our  numerical studies of the minima of the
VB  and SK  models. We  begin by  describing how  we found  the minima
numerically.   They  are   basically  just   quenches  from   infinite
temperature. In  Sec.  \ref{sec:overlap}  we have studied  the overlap
between the  minima and  we find  that the  minima produced  have only
trivial  overlaps  with  one  another. In  Sec  \ref{sec:marginal}  we
describe our  evidence that the minima  of the SK model  in zero field
have marginal stability  as they have an energy per  spin close to the
energy $E_c$  which marks the energy  at which the minima  starting to
have overlaps showing replica symmetry breaking features.

At zero temperature, the metastable states (minima) which we study are those obtained by aligning every spin along its  local field direction, starting  off from a random initial state. In the notation used for our numerical work based on Eq. (\ref{Ham}) we iterate the equations
\begin{equation}
\mathbf{S}^{n+1}_i= \frac{\mathbf{H}^{n}_i}{|\mathbf{H}_i^{n}|},
\label{eq:parn}
\end{equation}
where the local fields after the $n$th iteration, $\mathbf{H}_i^{n}$, are given by
\begin{equation}
\mathbf{H}_i^n= \sqrt{m} \mathbf{h}_i+m \sum_j J_{ij} \mathbf{S}_j^{n}.
\label{eq:hdefn}
\end{equation}
 For a given disorder
sample, a random configuration of spins is first created which would be a possible spin configuration at infinite temperature. Starting from the
first spin and scanning sequentially all the way up to the $N^{th}$
spin, every spin is aligned to its local field according to
Eq.~(\ref{eq:parn}), this whole process constituting one
sweep. The vector $(\Delta \mathbf{S}_{1},\Delta \mathbf{S}_{2},\cdots,\Delta \mathbf{S}_{N})$ is computed by subtracting the
spin configuration before the sweep from the spin configuration
generated after the sweep. The quantity $\eta = \frac{1}{Nm}\sum_{\mu=1}^{m}\sqrt{\sum_{j=1}^{N}(\Delta S_{j\mu})^{2}}$ is a measure of how close the configurations before and
after the sweep are. The spin configurations are iterated over many sweeps until the value of $\eta$ falls below $0.00001$, when the system is deemed to have converged to the metastable state described by Eq.~(\ref{eq:par}), which will be a minimum of the energy at zero temperature. Differing starting configurations usually generate different minima, at least for large systems.

\subsection{Overlap distribution}
\label{sec:overlap}
\begin{figure}
  \includegraphics[width=\columnwidth]{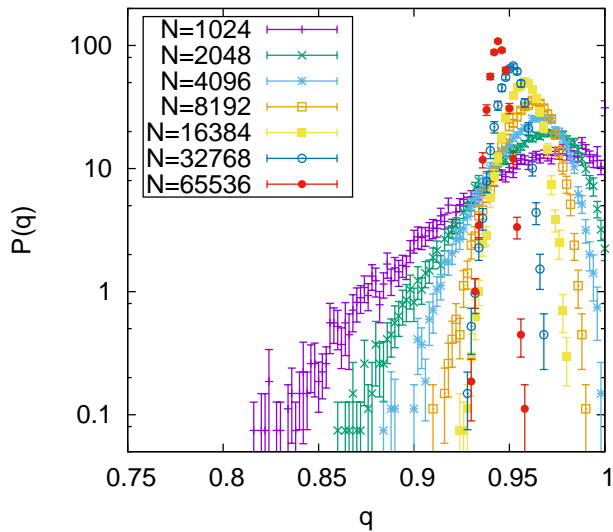} 
  \caption{(Color online) The overlap distribution $P(q)$ for the VB model ($\sigma =0, z =6$, $h_r=0.6$) for the minima generated by the prescription described in the text. $P(q)$ seems to be approaching a delta function as $N$ tends to infinity.}
  \label{fig0}
\end{figure}
It is informative to study the overlaps between the various minima.
Consider the overlap between two minima $A$ and $B$ defined as
\begin{equation}
q \equiv \frac{1}{N}\sum_{i}\mathbf{S}_i^{A} \cdot \mathbf{S}_i^{B}.
\end{equation}
Numerically, the following procedure is adopted. A particular realization of the bonds and fields is chosen. Choosing a random initial spin configuration, the above algorithm is implemented and descends to a locally stable state. This generates a metastable spin state that is stored. One then chooses a second initial condition, and the algorithm is applied, which generates a second metastable spin state which is also stored. One repeats this $N_{min}$ times generating in total $N_{min}$ metastable states (some or all of which might be identical). One then overlaps all pairs of these states, so there are $N_{pairs} = N_{min}(N_{min}-1)/2$ overlaps which are all used to make a histogram. The whole process is averaged over $N_{samp}$ samples of disorder. Fig.~\ref{fig0} shows the overlap distribution of the metastable states obtained by the above prescription for the VB model.  The figure suggests that in the thermodynamic limit, the distribution of overlaps, $P(q) = \delta(q-q_0(h_r))$. In zero field we have 
found that $q_0(h_r=0)=0$. Since we study  only a finite system of $N$ spins, the delta function peak is broadened to a Gaussian centered around $q_0$ and of width $O(\frac{1}{\sqrt{N}})$. We studied also the SK model, for a range of values for the $h_r$ fields, and the data are consistent with $P(q)$ just having a single peak in the thermodynamic limit. This suggests that the metastable states generated by the procedure of repeatedly putting spins parallel  to their local fields starting from a random state always produces minima which have a  $P(q)$ of the same type as would be expected for the paramagnetic phase.

 Newman and Stein \cite{newman:99} showed that for Ising spins in zero field that when one starts off from an initial state, equivalent to being at infinite temperature, and quenches to zero temperature one always ends up in a state with a trivial $P(q)=\delta(q)$, in agreement, for example with the study of Parisi \cite{parisi:95}. Our results for vector spin glasses seem exactly analogous to the Ising results.
 
\subsection{Marginal stability}
\label{sec:marginal}
In  this subsection  we  shall focus  on  the  Ising, XY  ($m  = 2$)  and
Heisenberg ($m=3$) SK models with zero random field.  Parisi found for
the Ising case that when  starting a quench from infinite temperature,
when  the spins  are  just  randomly up  or  down,  and putting  spins
parallel to  their local fields  according to various  algorithms, the
final   state    had   an   energy   per    spin   $\varepsilon=-0.73$
\cite{parisi:95}.  In  their  studies  of one-spin  flip  stable  spin
glasses in zero field, Bray  and Moore  \cite{bm1981,bm:81a} found that  such states
associated with  a trivial $P(q)=\delta(q)$  should not exist  below a
critical energy $E_c$ and for the Ising case $E_c=-0.672$. States with
an  energy close  to $-0.73$  would be  expected to  be have  a $P(q)$
rather similar to those for  full replica symmetry breaking, but those
generated in the quench have a  trivial $P(q)$. There is no paradox as
the  states generated  in  the  quench have  more  than one-spin  flip
stability  \cite{yan:15}.  This  results  in a  distribution of  local
fields behaving  at small  fields so that $p(h)  \sim h$,  very different
from  that expected  from the  study of  the $p(h)$  of one-spin  flip
stable  states  \cite{roberts:81}  for  which $p(0)$  is  finite,  and
instead similar to what is found in the true ground state -- the state
which is stable  against flipping an arbitrary number of  spins. It is
by that  means that the  theorem of Newman and  Stein \cite{newman:99}
that in a  quench from a random initial state  the final $P(q)$ should
be  trivial is  realized, despite  the  quenched energy  being in  the
region  where one  would expect  the $P(q)$  of one  spin flip  stable
states to be non-trivial. The change  in the form of $p(h)$ means that
the true $E_c$ is  not at $-0.672$, but instead is  at least closer to
$-0.73$.

\begin{figure}
  \includegraphics[width=\columnwidth]{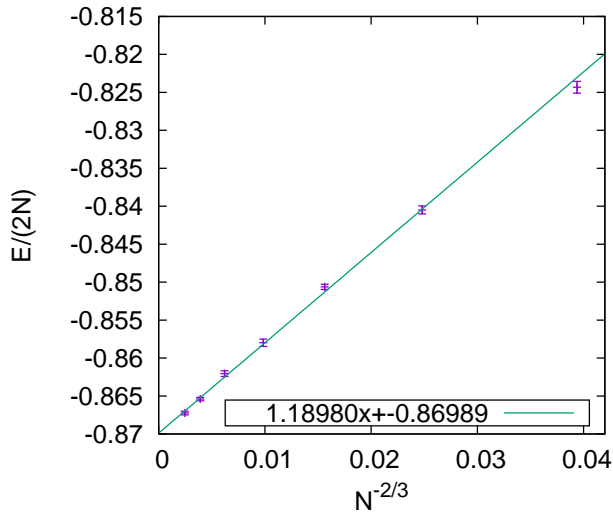} 
  \caption{(Color online) The average  energy per site and spin component for the XY SK spin glass model ($m =2$) with $h_r=0$ plotted against $1/N^{2/3}$ in order to estimate the infinite system value of the energy obtained from a quench from infinite temperature. For $m=2$, $E_c=-0.866$ \cite{bm1981}.}
  \label{fig:SKXYEc}
\end{figure}

\begin{figure}
  \includegraphics[width=\columnwidth]{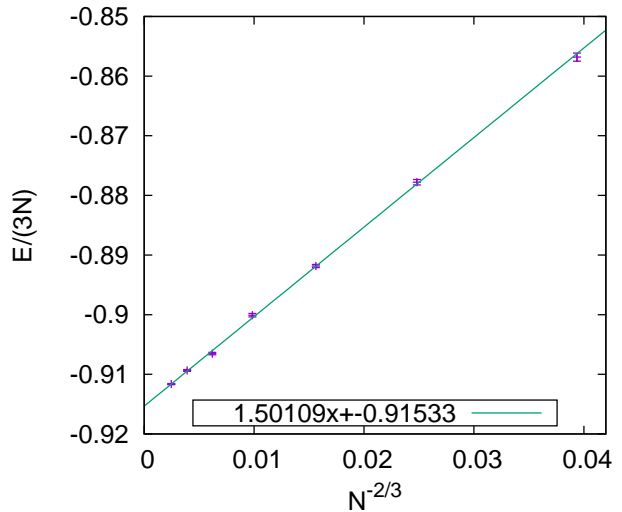} 
  \caption{(Color online) The average energy per site and spin component for the Heisenberg SK spin glass ($m =3$) with $h_r=0$ plotted against $1/N^{2/3}$ in order to estimate the infinite system value of the energy obtained from a quench from infinite temperature. For $m =3$, $E_c=-0.914$ \cite{bm1981}. }
  \label{fig:SKHeisEc}
\end{figure}

For the vector SK spin glasses in zero field we have studied the
energy reached in a quench from infinite temperature by putting the
spins parallel to their local fields.  In Figs.  \ref{fig:SKXYEc} and
\ref{fig:SKHeisEc} we have plotted our estimates of this energy as a
function of $1/N^{2/3}$, the form commonly used for the energy size
dependence of the SK model \cite{boettcher:03, billoire:08}.  For
$m=2$, the extrapolated energy per spin component is $\approx -0.870$,
whereas its $E_c=-0.866$ according to the analysis in
Ref. \onlinecite{bm1981}; for $m=3$ the extrapolated energy per spin
component is $\approx -0.915$ whereas its $E_c =-0.914$ \cite{bm1981}.
Minima whose energies lie below the critical energy $E_c$, are
associated with non-trivial (i.e.  RSB) form for their $P(q)$,
calculated from the overlaps of the minima at the same energy
\cite{bm1981, bm:81a}.  We found just as for the Ising SK model that
the energy reached in the quench varied little when the greedy
algorithm was used instead of the sequential algorithm
\cite{parisi:95}.

As  the energy  of  the  quenched state  is  remarkably  close to  the
critical energies  calculated by  Bray and  Moore \cite{bm1981,bm:81a}
for  $m=2$ and  $m=3$, this  suggests that  the state  reached in  the
quench     is     well-described     by    the     calculations     in
Ref. \onlinecite{bm1981},  whereas for  the Ising case  the equivalent
calculation which enumerates the number of one-spin flip stable states
does not  give the resulting $p(h)$  of the quenched states  with much
accuracy and so does not produce an accurate estimate of $E_c$.

One knows a  lot about behavior at  $E_c$ at least for  Ising spins in
zero  random  field  \cite{bm:81a}.  For states  of  energy  per  spin
$\varepsilon >  E_c$, the  annealed and  quenched averages  agree with
each other, but for energies $\varepsilon < E_c$, the two calculations
differ.  As  $\varepsilon$  approaches  $E_c$, behavior  is  as  at  a
critical point,  with growing  length scales  etc. and  massless modes
\cite{bm:81a}. For the  Ising case the properties of  these modes were
discussed in  Ref. \onlinecite{bm:81a}.  We intend  to return  to this
topic in a future publication for the case of vector spin glasses.

 When one sets an Ising spin parallel to its local field in the course
 of the  quench, spin avalanches  may be  triggered. If the  number of
 neighbors $z$ is of order $N$ then  the avalanches can be on all size
 scales \cite{boettcher:08, andresen:13}.  Thus  the Ising SK model is
 an  example of  a  system  with marginal  stability  as discussed  by
 M\"{u}ller   and   Wyart   \cite{muller:15}.   It   was   argued   in
 Ref. \onlinecite{muller:15} that as  the quench progresses the system
 will reach the marginal manifold which separates stable from unstable
 configurations. As  this point is  approached the dynamics  slows and
 eventually freezes near the marginal  manifold.  The VB model with $z
 =6$ does not have large  scale avalanches \cite{andresen:13} and does
 not have any marginal features; a first study of avalanches in the undiluted
 one-dimensional long-range  models can be found in \cite{boettcher:08}.   While the  Ising VB model  does not have  large scale
 avalanches, there certainly  will be an energy $E_c$  below which the
 minima will have  non-trivial overlaps. What is not  clear is whether
 it is the large  avalanches which ensures that the states  generated in a
 quench are close to this energy.
 
We  also  do not  know  what  difference  the  existence of  a  finite
temperature  phase  might  make  to the  properties  of  the  quenched
state. For example,  are there features of the quenched  states of one
and  two dimensional  Ising spin  glasses,  where there  is no  finite
temperature  spin glass  transition, which  differ significantly  from
those of  the three dimensional  spin glass,  where there is  a finite
temperature phase transition? We also  do not know what features might
arise if  there is  a phase  transition to a  state with  full replica
symmetry breaking, as opposed to a state with just replica symmetry.

For systems  for which the  excitations are  not discrete, such  as in
vector spin  glasses, marginality  takes a  different form,  and seems
related  to the  development of  negative eigenvalues  in the  Hessian
\cite{muller:15,sharma2014avalanches}.  Such  eigenvalue instabilities
might be triggered in a quench  where one puts spins parallel to their
local fields. On the other hand,  one could imagine a steepest descent
procedure  starting from  the  initial spin  orientation and  smoothly
proceeding  to a  minimum. Does  that result  in a  final state  whose
properties differ  from those generated  by putting spins  parallel to
their local fields? 

There are many topics which should be studied! We believe that the proximity of the quenched energy to the calculated critical energy $E_c$, at least for the cases of $m =2$ and $m =3$  will provide valuable analytical insights concerning marginal stability. One of our motivations for the analytic work in the next section was to calculate $E_c(h_r)$ in the presence of a non-zero random vector field, but, as we shall see, algebraic difficulties prevented us from achieving this goal. But it would be good to know how general is the result that the energy obtained in a quench coincides with the energy at which the overlaps of the minima display replica symmetry breaking features. 

\section{Metastable states in the SK model in the presence of a random field}
\label{sec:SKanalytic}

In this section we follow the method of Ref. \onlinecite{bm1981}
 to study the complexity and Hessian properties of the minima for the SK model but in the presence of a random vector field. We begin by writing down the first steps in the formalism following Ref. \onlinecite{bm1981}. In
 subsection \ref{SKannealed} we show that within the annealed approximation, where one averages $N_S(\varepsilon)$ itself over the bonds $J_{ij}$ and the random fields $h_i^{ex}$ analytical progress is fairly straightforward. Fortunately the annealed approximation is also exact for fields $h_r > h_{AT}$. In subsection \ref{sec:quenched} we describe our attempts to solve the quenched case. We believe that our approach based on  replica symmetry assumptions should be good down to its limit of stability which would be at $E_c(h_r)$, but  algebraic difficulties prevented us from actually determining $E_c(h_r)$.

 We  find it convenient to write the Hamiltonian for the $m$-vector spin glass in an $m$-component external field  as
\begin{equation}
 \mathcal{H}=-\frac{m}{2}\sum_{i,j} J_{ij}\bm{S}_i\cdot\bm{S}_j-m\sum_i\bm{h}^{\rm ex}_i\cdot\bm{S}_i,
 \label{Hamil}
\end{equation}
where the $m$-component spins $\bm{S}_i=\{S^\alpha_i\}$, ($\alpha=1,\cdots,m$, $i=1,\cdots,N$) have a unit length
$S_i=1$. The interactions $J_{ij}$ are chosen from a Gaussian distribution with zero mean and the variance $1/N$. 
In this section, for convenience, we use the notation $\bm{h}^{\rm ex}_i=\bm{h}_i/\sqrt{m}$ for
the random Gaussian external fields with zero mean and the variance  
\begin{equation}
 \langle h^{{\rm ex},\alpha}_i h^{{\rm ex},\beta}_j\rangle =\frac{h^2_r}{m}\delta_{ij}\delta^{\alpha\beta}.
\end{equation}

At zero temperature, the spins are aligned in the direction of the local internal field $\bm{H}_i$, i.e.
\begin{equation}
 \bm{S}_i=\hat{\bm{H}}_i\equiv\frac{\bm{H}_i}{H_i},
\label{eq:par}
\end{equation}
where
\begin{equation}
 \bm{H}_i=\sum_j J_{ij}\bm{S}_j +\bm{h}^{\rm ex}_i.
\label{eq:hdef}
\end{equation}
In terms of the local fields, the ground state energy $E$ can be written as
\begin{equation}
 E=-\frac{m}{2}\sum_i(H_i+\hat{\bm{H}}_i\cdot \bm{h}^{\rm ex}_i). 
\end{equation}

The number of metastable states with energy $\varepsilon$ per site and per spin component is given by
\begin{align}
 &N_S(\varepsilon)=\int\prod_{i,\alpha}dH_i^{\alpha}\int\prod_{i,\alpha}dS_i^{\alpha} 
\prod_{i,\alpha}\delta(S_i^{\alpha}-\hat{H}_i^{\alpha}) \nonumber \\
&~~~~~~\times \prod_{i,\alpha}\delta\left(H_i^{\alpha}-\sum_j J_{ij}S_j^{\alpha}-h^{{\rm ex},\alpha}_{i}\right)
|\det\mathsf{M}\{J_{ij}\}|
 \nonumber \\
 &~~~~~~\times
 \delta\left(Nm\varepsilon+\frac 1 2 m \sum_i (H_i + \hat{\bm{H}}_i\cdot\bm{h}^{\rm ex}_i ) \right),
 \label{NS}
\end{align}
where 
\begin{equation}
 M^{\alpha\beta}_{ij}=\frac{\partial}{\partial S^\beta_j}(S^\alpha_i-\hat{H}^\alpha_i)
 =\delta_{ij}\delta^{\alpha\beta}-J_{ij}\frac{P^{\alpha\beta}_i}{H_i}
\end{equation}
with $P^{\alpha\beta}_i\equiv \delta^{\alpha\beta}-\hat{H}^\alpha_i\hat{H}^\beta_i$
is the projection matrix.

\subsection{Annealed Approximation}
\label{SKannealed}
We now calculate the average of $N_S(\varepsilon)$ over the random couplings and the random external fields. As we will 
see below, the direct evaluation of the quenched average $\langle \ln N_S(\varepsilon) \rangle$ is very complicated. 
Here we first present the annealed approximation, where we evaluate the annealed complexity $g_A(\varepsilon)=\ln \langle N_S(\varepsilon) \rangle/N$.
The whole calculation is very similar to those in Appendix 2 of Ref.~\onlinecite{bm1981} except for 
the part involving the average over the random field. Below we sketch the calculation.

The first delta functions in Eq.~(\ref{NS}) can be integrated away. We use the integral
representations for the second and third delta functions using the variables $x_i^{\alpha}$ and $u$, respectively,
along the imaginary axis. 
The average over the random couplings can be done in an exactly the same way as in Ref.~\onlinecite{bm1981}.
We briefly summarize the results below. The random couplings appear in the factor 
\begin{align}
& \left\langle 
 \exp\Big[-\sum_{i<j}J_{ij}\sum_{i,\alpha} (x^\alpha_{i}\hat{H}^\alpha_{j}
+x^\alpha_{j}\hat{H}^\alpha_{i})\Big]  |\det\mathsf{M} \{J_{ij}\}| 
 \right\rangle_J \nonumber \\
=&\exp \Big[ \frac 1 {2N} \sum_{i<j}\Big\{ \sum_{\alpha}(x^\alpha_{i}\hat{H}^\alpha_{j}
+x^\alpha_{j}\hat{H}^\alpha_{i}) \Big\}^2\Big] \nonumber \\ 
&~~~~~~~~~~~~~\times \left\langle  |\det\mathsf{M} \{J_{ij}-O(\frac 1 N)\}|  \right\rangle_J .
 \end{align}
After neglecting the $O(1/N)$ term, we evaluate the average of the determinant as \cite{bm1981}
\begin{equation}
\left\langle  |\det\mathsf{M} \{J_{ij}\}|  
\right\rangle_J=\exp(\frac 1 2 Nm\bar{\chi})\prod_i \left(1-\frac{\bar{\chi}}{H_i}\right)^{m-1},
\label{det}
\end{equation}
where the susceptibility $\bar{\chi}$ satisfies the self-consistency equation \cite{bm1981}
\begin{equation}
\bar{\chi}=(1-\frac 1 m)\frac 1 N \sum_i \frac 1 {H_i-\bar{\chi}}
\label{chi}
\end{equation}
with the condition $H_i\ge\bar{\chi}$.
Using the rotational invariance and the Hubbard-Stratonovich transformation, 
we can rewrite the exponential factor in front of the determinant
as
\begin{align}
 &\exp[\frac 1{2m}\sum_{i,\alpha}(x^\alpha_{i})^2 ]  \\
&~\times \int\frac{dv}{(2\pi/Nm)^{1/2}}\;
 \exp [ -\frac{Nm}{2} v^2+ v 
 \sum_{i,\alpha}x^\alpha_{i}\hat{H}^\alpha_{i} ]. \nonumber
\end{align}
In the present case, we have to    
average over the random field. Collecting the relevant terms, we have
\begin{align}
& \left\langle 
 \exp\Big[-\sum_{i,\alpha} (x^\alpha_{i}+\frac 1 2 u m \hat{H}^\alpha_{i})h^{{\rm ex},\alpha}_{i}\Big]
 \right\rangle_{\bm{h}^{\rm ex}}  \\
=&\exp\Big[ 
\frac{h^2_r}{2m}\sum_{i,\alpha}(x^\alpha_{i})^2
+\frac{h^2_r}{2} u  \sum_{i,\alpha}x^\alpha_{i}\hat{H}^\alpha_{i}
+Nm\frac{h^2_r }{8}u^2
\Big] .   \nonumber  
 \end{align}

All the site indices are now decoupled. We express the condition Eq.~(\ref{chi})
using the integral representation of the delta function with the variable $\lambda$
running along the imaginary axis. 
Putting all the terms together, we have 
\begin{align}
 &\langle [N_S(\varepsilon)] \rangle_{J,h^{\rm ex}}=
 \int \frac{du}{2\pi i} \int \frac{dv}{\sqrt{2\pi/Nm}} \int d\bar{\chi} 
 \int \frac{d\lambda}{2\pi i}  \nonumber  \\
& \times
\exp\Big[
Nm \lambda \bar{\chi} + \frac{Nm}{2}  \bar{\chi}^2 
-Nm\varepsilon  u 
-\frac{Nm}{2} v^2  \nonumber \\
&~~~~~~~~~~~~~~~~~~~~~~~+Nm\frac{h^2_r}{8}u^2 +N\ln I^\prime 
\Big],  \label{NS_ann}
 \end{align}
where
\begin{align}
 I^\prime=&\int_{H\ge\bar{\chi}}\prod_{\alpha}dH^\alpha
 \int\prod_{\alpha}\frac{dx^\alpha}{2\pi i} 
 \left(1-\frac{\bar{\chi}}{H}\right)^{m-1}  \nonumber \\
 &\times \exp\Bigg[
 \frac{1+h^2_r}{2m}\sum_{\alpha}(x^\alpha)^2
+ (v+\frac{h^2_r}{2}u)\sum_\alpha x^\alpha\hat{H}^\alpha 
      \nonumber \\
 & +\sum_{\alpha}x^\alpha H^\alpha -(m-1) \lambda (H-\bar{\chi})^{-1}
 -\frac{m}{2} u H 
 \Bigg]  
  \end{align}
The Gaussian integral over $x^\alpha$ can be done analytically.  
The integrals in Eq.~(\ref{NS_ann}) are evaluated via the saddle point method in the $N\to\infty$ limit.
Following the procedure described in Ref.~\onlinecite{bm1981},
we introduce new variables 
$\bm{h}\equiv (H-\bar{\chi})\hat{\bm{H}}$ and
$\Delta=-v-\bar{\chi}$ and use the saddle point condition for $\bar{\chi}$, which is
\begin{equation}
\lambda-\Delta-\frac u 2=0.
\end{equation}

We finally have an expression for the annealed complexity 
$g_A(\varepsilon)\equiv N^{-1}\ln\langle N_S(\varepsilon)\rangle$ as
\begin{equation}
 g_A(\varepsilon) = m (-\frac{\Delta^2}{2} 
-\varepsilon  u +\frac{h^2_r}{8}u^2) +\ln I, 
 \end{equation}
 where 
 \begin{align}
 &I= \left(\frac{m}{2\pi(1+h^2_r)}\right)^{m/2}S_m\int^\infty_0 dh\; h^{m-1} 
   \\
 &\times\exp\Big[ 
 -\frac{m}{2(1+h^2_r)}(h-\Delta+\frac{h^2_r}{2}u)^2 \nonumber \\ 
&~~~~~~~~~~~~~~~ -\frac{(m-1)}{h}(\Delta+\frac u 2)
 -\frac{m}{2}uh\Big] \nonumber
\end{align}
with the surface area of the $m$-dimensional unit sphere $S_m=2\pi^{m/2}/\Gamma(m/2)$.
The parameters $\Delta$ and $u$ are determined variationally as 
$\partial g_A/\partial \Delta =\partial g_A/\partial u =0$.

We focus on the total number of metastable states, which are obtained by integrating 
$\exp(Ng_A(\varepsilon))$ over $\varepsilon$, or equivalently by setting $u=0$. Thus we are effectively focussing on the most numerous states, those at the top of the band where $g_A(\varepsilon)$ is largest. 
In this case, $g_A=-(m/2)\Delta^2+\ln I_0$, where
\begin{align}
 I_0 =& S_m \left(\frac{m}{2\pi(1+h^2_r)}\right)^{m/2}\int_0^\infty dh\; h^{m-1} \nonumber \\
&\times \exp\left[ -(m-1)\frac{\Delta}{h}-\frac{m(h-\Delta)^2}{2(1+h^2_r)}\right]
\label{I0}.
\end{align}
The parameter $\Delta$ is determined by the saddle point equation 
\begin{equation}
 \Delta=\frac{1}{2+h^2_r}\langle h \rangle -
 \left(1-\frac{1}{m}\right)\left(\frac{1+h^2_r}{2+h^2_r}\right)
 \left\langle\frac{1}{h}\right\rangle,
 \label{Delta}
\end{equation}
where the average is calculated with respect to the probability distribution for the internal field given by
the integrand of 
$I_0$ in Eq.~(\ref{I0}).
Using $\langle h \rangle =\Delta + \langle h-\Delta \rangle$, we can rewrite Eq.~(\ref{I0}) as
\begin{equation}
 \Delta \left[ 1- \left(1-\frac 1 m \right)\left\langle \frac 1 {h^2} \right\rangle \right]=0.
\label{Delta1}
 \end{equation}

For various values of the external field $h_r$,
we solve numerically Eq.~(\ref{Delta}). For $m=3$, we find that when $h_r>h_{AT}=1$
there is only a trivial solution, $\Delta=0$. (Note that the Almeida-Thouless field $h_{AT}$ at $T=0$ is
$h_{AT}=1/\sqrt{m-2}$ ~\cite{sharma2010almeida}). From Eq.~(\ref{I0}), we see that in this case $I_0=1$ and the complexity $g$ vanishes above the AT field. 
For $h_r<h_{AT}$, a nontrivial solution, $\Delta\neq 0$ exists. 
We find that the values of $\Delta$ and $g_A$ increase as the external field $h_r$ decreases
from $h_{AT}$, and approach the known values, 0.170 and 0.00839 at zero external field \cite{bm1981}.
For $h_r$ smaller than but very close to $h_{AT}$, $\Delta$ is very small. 
We may obtain an analytic expression for $g_A$ in this case. 
By expanding everything in Eq.~(\ref{Delta1}) in powers of $\Delta$, we find for $m=3$ that 
\begin{equation}
 g_A=\frac 3 2 (h^2_{AT}-h^2_r)\tilde{\Delta}^2+8\sqrt{\frac{3}{2\pi}}
 \tilde{\Delta}^3\ln\tilde{\Delta}+O(\tilde{\Delta}^3),
 \label{gA0}
\end{equation}
where $\tilde{\Delta}=\Delta/\sqrt{1+h^2_r}$. The fact that $g_A$ must be stationary with respect to $\tilde{\Delta}$, enables one to determine how the complexity vanishes as $h_r \to h_{AT}$ and the value of $\tilde{\Delta}$ in this limit.

Using the distribution for the internal field $H$ (or $h$), 
we first calculate the spin glass susceptibility $\chi_{SG}\equiv (Nm)^{-1}\mathrm{Tr} \mathbf{\chi}^2$ with the 
susceptibility matrix $\mathbf{\chi}=\chi_{ij}^{\alpha\beta}$ \cite{bm1981}.  
Note that the susceptibility in Eq.~(\ref{chi}) is
just $\bar{\chi}=(Nm)^{-1}\mathrm{Tr}\mathbf{\chi}$. The spin glass susceptibility is given by \cite{bm1981}
$\chi_{SG}=(1-\lambda_R)/\lambda_R$, where
\begin{equation}
\lambda_R=1-(1-\frac 1 m)\frac 1 N \sum_i \frac 1{(H_i-\bar{\chi})^2}.
\label{lambda_R}
\end{equation}
This quantity is exactly the one in the square bracket in Eq.~(\ref{Delta1}). 
Therefore, since $\Delta\neq 0$ for $h_r<h_{AT}$,
$\lambda_R$ vanishes and consequently $\chi_{SG}$ diverges. Above the AT field,
there is only a trivial solution $\Delta=0$. In this case the integrals are just Gaussians and we can evaluate explicitly $\frac{1}{N} \sum_i \frac 1{(H_i-\bar{\chi})^2}$, with the result that $\lambda_R=(h_r^2-1/(m-2))/(1+h_r^2)$, so 
 the spin glass 
susceptibility as a function of the external random field for $h_r>h_{AT}$ is given by
\begin{equation}
\chi_{SG}=\frac{1+h_{AT}^2}{h_r^2-h_{AT}^2},
\label{chisgexact}
\end{equation}
provided $h_r > h_{AT}$ and $m > 2$. The simple divergence of $\chi_{SG}$ as $h_r \to h_{AT}$ is a feature of the SK limit and is not found in the Viana-Bray model at least amongst the quenched states of our numerical studies, see Sec. \ref{sec:spinglasssusceptibility}

We now calculate the eigenvalue spectrum of 
the Hessian matrix $\mathsf{A}$. 
The calculation closely follows the steps in Ref.~\onlinecite{bm1982} for the case of zero external field.  
We consider (transverse) fluctuations around the $T=0$ solution $\bm{S}^{0}_i\equiv\hat{\bm{H}}_i$ by writing
$\bm{S}_i=\bm{S}^{0}_i+\bm{\epsilon}_i$, where $\bm{\epsilon}_i=\sum_\alpha \epsilon^\alpha_i \hat{\bm{e}}_\alpha(i)$
with the $(m-1)$ orthonormal vectors $\hat{e}_\alpha(i)$, $\alpha=1,\cdots,m-1$ satisfying
$\bm{S}^{0}_i\cdot \hat{\bm{e}}_\alpha(i)=0$. Inserting this into Eq.~(\ref{Hamil}), we have the Hessian matrix
as
\begin{equation}
A^{\alpha\beta}_{ij}\equiv\frac{\partial(\mathcal{H}/m)}{\partial\epsilon^\alpha_i \partial\epsilon^\beta_j}
=H_i\delta_{ij}\delta^{\alpha\beta}-J_{ij} \hat{\bm{e}}_\alpha(i)\cdot\hat{\bm{e}}_\beta(j).
\end{equation}
The eigenvalue spectrum $\rho(\lambda)$ can be calculated from the resolvent 
$\mathsf{G}=(\lambda\mathsf{I}-\mathsf{A})^{-1}$ as
\begin{equation}
\rho(\lambda)=\frac{1}{N(m-1)\pi}\mathrm{Im} \; \mathrm{Tr} \mathsf{G}(\lambda-i\delta),
\label{eqn:rho}
\end{equation}
where $\mathsf{I}$ is the $(m-1)N$-dimensional unit matrix and $\delta$ is an infinitesimal positive number.
The locator expansion method \cite{bm1979} is used to evaluate $\rho(\lambda)$, which  yields the following
self-consistent equation for $\bar{G}(\lambda)\equiv ((m-1)N)^{-1} \mathrm{Tr} \mathsf{G}(\lambda)$:
\begin{equation}
\bar{G}(\lambda)=\left\langle \frac 1{\lambda-H-(1-\frac 1 m)\bar{G}(\lambda)}\right\rangle,
\label{Gbar}
\end{equation}
where $\langle ~\rangle$ denotes the average over the distribution for $h$ given in the integrand in Eq.~(\ref{I0}). 
Note that $H=h+\bar{\chi}$ and $\bar{\chi}=(1-1/m)\langle 1/h\rangle$ from Eq.~(\ref{chi}).
We first separate $\bar{G}=\bar{G}^\prime+i\bar{G}^{\prime\prime}$ into real and imaginary parts and solve
Eq.~(\ref{Gbar}) numerically for $\bar{G}^\prime(\lambda)$ 
and $\bar{G}^{\prime\prime}(\lambda)$ as a function of $\lambda$. 
The eigenvalue spectrum is just $\rho(\lambda)=\pi^{-1}\bar{G}^{\prime\prime}(\lambda)$.

\begin{figure}
  \includegraphics[width=\columnwidth]{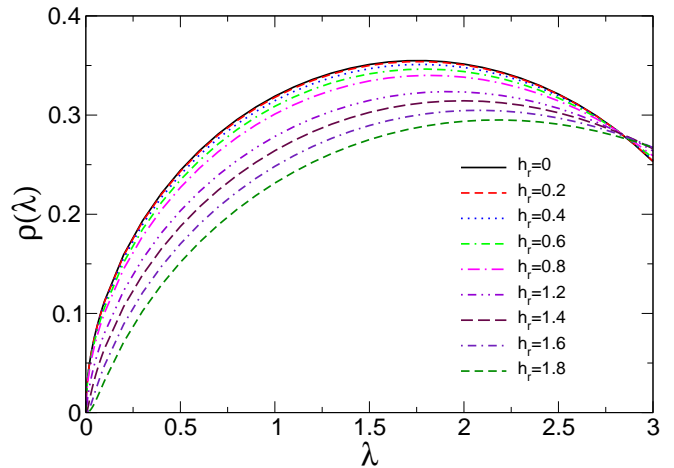} 
  \caption{(Color online) The eigenvalue spectrum of the Hessian at zero temperature 
  for the vector spin glass with $m=3$ in the SK limit.The various lines correspond to different values of  
  $h_r$,  the external random field.}
  \label{fig:rho}
\end{figure}
\vspace{0.2cm}

\begin{figure}
  \includegraphics[width=\columnwidth]{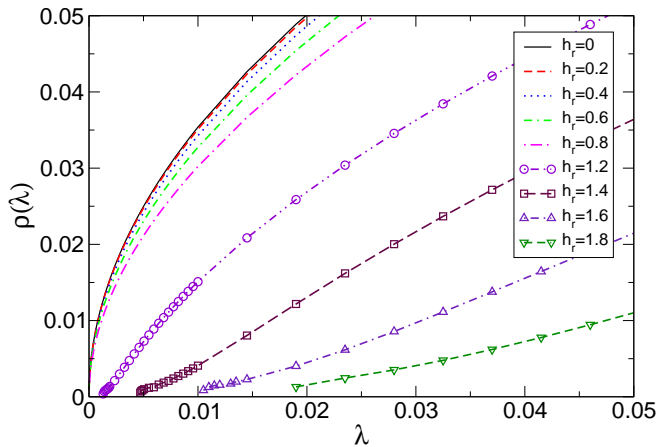} 
  \caption{(Color online) The magnified view of the same figure as Fig.~\ref{fig:rho} but for the small eigenvalues.}
  \label{fig:rho_mag}
\end{figure}

As we can see from Figs.~\ref{fig:rho} and \ref{fig:rho_mag}, $\rho(\lambda)$ does not change very much as 
we increase $h_r$ from zero up to $h_{AT}=1$. 
For the external field larger than the AT field, however, Fig.~\ref{fig:rho_mag}\
clearly shows that the eigenvalue spectrum develops a gap. The gap increases with
the increasing external field. By directly working on Eq.~(\ref{Gbar}) in 
the small-$\lambda$ limit, we find that for small eigenvalues
\begin{equation}
 \rho(\lambda)\simeq \frac{1}{\pi(1-1/m)}\frac{1}{\sqrt{s}}\sqrt{\lambda-\lambda_0},
\end{equation}
where $s=(1-m^{-1})\langle 1/h^3 \rangle$ and $\lambda_0=\lambda^2_R/4s$ with $\lambda_R$ defined in Eq.~(\ref{lambda_R}).
Our numerical solution of the equations for $G(\lambda)$ confirms that there is no gap below $h_{AT}$ which is consistent with the previous observation that $\lambda_R$ vanishes
there. However,  the integral by which $s$ is defined diverges for $h_r > h_{AT}$ when $m <3$ and we no longer see a square root singularity at the band-edge. In the case of $m =3$ our numerical solution shown in Fig. \ref{fig:rho_mag} suggests instead of the square root dependence there is a roughly linear dependence as $\lambda$ approaches the numerically determined band-edge $\lambda_0$, but unfortunately we have not been able to derive its form analytically. Fig. \ref{fig:rho} shows that away from $\lambda_0$ the density of states is rather as if it had the square root form. As $h_r \to h_{AT}$ this square root form works all the way to zero.

\subsection{Quenched Average}
\label{sec:quenched}
In  this subsection, we  attempt to  evaluate the  quenched complexity
$g(\varepsilon)=N^{-1}\langle\ln     N_S(\varepsilon)\rangle$.     The
calculations  are  quite  complicated  and  some of  the  details  are
sketched  in  the  Appendix.   In  order  to  calculate  $\langle  \ln
N_S(\varepsilon) \rangle$,  we consider  an average of  the replicated
quantity  $\langle [N_S(\varepsilon)]^n  \rangle_{J,h^{\rm  ex}}$.  We
then  have  an expression  similar  to  Eq.~(\ref{NS_ann}), where  the
integrals  are  now  over  replicated variables,  $u^\eta$,  $v^\eta$,
$\bar{\chi}^\eta$   and  $\lambda^\eta$   with  the   replica  indices
$\eta,\mu=1,\cdots,n$.   In  addition to  these,  the expression  also
involves  the  integrals  over  the  variables  carrying  off-diagonal
replica indices, which  are denoted by $A_{\eta\nu}$, $A^*_{\eta\nu}$,
$B_{\eta\nu}$ and $B^*_{\eta\nu}$ with  $\eta<\nu$.  In the absence of
external    field,    it    can    be   shown    \cite{bm1981}    that
$A_{\eta\nu}=A^*_{\eta\nu}  =B_{\eta\nu}=B^*_{\eta\nu}=0$ is  always a
solution to the saddle point equations.   It is shown to be stable for
$\varepsilon> E_c$  for the  $E_c$, for  which the
quenched average  coincides with the  annealed one.  For  $h_r\neq 0$,
however,   we    find   that   this    is   no   longer    the   case.
$A_{\eta\nu}=A^*_{\eta\nu}=B_{\eta\nu}=B^*_{\eta\nu}=0$   is   not   a
solution  to  saddle  point  equations.  The  saddle  point  solutions
involve  nonvanishing off-diagonal variables  in replica  indices.  We
find that in general the saddle point equations are too complicated to
allow  explicit  solutions.  (See  the  Appendix  for details.) 

  The
 quenched  average  is    different from  the
annealed  one  for  a  finite  external field   when  $h_r<
h_{AT}$.  When $h_r> h_{AT}$  the annealed  and quenched  averages are
identical  in  every  way  for  the  SK  model,  which  has  vanishing
complexity in this region. We doubt whether the same statement is true
for  any  model  such  as  the Viana-Bray  model  which  has  non-zero
complexity for $h_r > h_{AT}$. We also do not know for sure whether our replica
symmetric solution for $A_{\eta \nu}$  etc. is stable. It is possible
that even  at $u=0$  there is a  need to  go to full  replica symmetry
breaking.  Unfortunately 
algebraic complexities  have prevented us from even finding a  solution of
the replica  symmetric equations,  so determining their  stability looks very challenging. However, the results of the numerical work reported on the form of $P(q)$ in Sec. \ref{sec:metastability} for the Viana-Bray model in a field suggests that the states reached in the quench have replica symmetry.

We look for the saddle points in the replica symmetric form,
\begin{align}
&A_{\eta\nu}=A,~~A^*_{\eta\nu}=A^*,~~B_{\eta\nu}=B^*_{\eta\nu}=B, \nonumber \\
&u^\eta=u,~~v^\eta=v,~~\bar{\chi}^\eta=\bar{\chi},~~\lambda^\eta=\lambda.
\end{align}
After a lengthy calculation (see Appendix), we arrive at the expression for the quenched complexity as follows.
\begin{align}
g(\varepsilon)=&
m \Big\{ -\frac{\Delta^2}{2} -\varepsilon u   - \frac{A}{2m}
+\frac{1}{2} (AA^*+B^2)\Big\} 
\label{geps}  \\
+&
\int \frac{d^m \bm{w}}{(2\pi)^{m/2}}
\int \frac{d^m \bm{y}}{(2\pi)^{m/2}}
\int \frac{d^m \bm{z} d^m\bm{z}^*}{(2\pi)^m}\; \nonumber \\
&\times \exp[-\frac 1 2 \sum^m_\alpha (w^2_\alpha+y^2_\alpha+z_\alpha z^*_\alpha)]
\; \ln K(\bm{w},\bm{y},\bm{z},\bm{z}^*), \nonumber
\end{align}
where
\begin{align}
 K=&
 \int d^m\bm{h}
 \int^{i\infty}_{-i\infty} \frac{d^m \bm{x}}{2\pi i} 
 \;\exp\Bigg[
 \frac{1-mA^*}{2m}\bm{x}^2 
 \nonumber \\
&+ (h-\Delta-B)\bm{x}\cdot\hat{\bm{h}}
  -(m-1)\frac{\Delta + u/2}{h}
 -\frac{m}{2}uh \nonumber \\
 &+
 \sqrt{A^*+\frac{h^2_r}{m}} \;\bm{w}\cdot\bm{x}
 +\sqrt{A+\frac{mh^2_r}{4}u^2 }  \; \bm{y}\cdot  \hat{\bm{h}} \nonumber \\
&+ \sqrt{\frac{1}{2}(B+\frac{h^2_r}{2}u) }
\left(\bm{z}\cdot\bm{x}+\bm{z}^* \cdot \hat{\bm{h}}\right)
 \Bigg].
\end{align}
All the parameters, $\Delta$, $A$, $A^*$, $B$ and $u$ are to be determined in a variational way.
We found, however, that it is very difficult to solve the saddle point equations 
and obtain the quenched complexity, even numerically. 

For the total number of metastable states, $u=0$, we can find a simple solution to saddle point equations at 
$\Delta=A=B=0$ and $A^*=1/m$. In this case, $K=1$ and the complexity $g$ vanishes. This solution 
must describe the case where $h_r>h_{AT}$ and it is identical to the annealed average.
For the external field $h_r$ just below $h_{AT}$, $\Delta$, $A$, $B$ and $C\equiv 1/m-A^*$
are expected to be very small, and we may expand the integrals in Eq.~(\ref{geps})
in these variables.
We find after a very lengthy calculation that
\begin{equation}
 g\simeq 
\frac{m}{1+h^2_r} (h^2_{AT}-h^2_r)\Big[  \frac{\Delta^2}{2}   + \frac{AC}{2}
 -  \frac{B^2}{2}  \Big] .
\end{equation}
Note that from Eq.~(\ref{saddle}), we expect $B$ is pure imaginary. 
In order to determine how these variables behave near $h_{AT}$, we need higher order terms.
Unfortunately, the complicated nature of these equations, however, has prevented us from 
going beyond the quadratic orders. It seems natural to expect that the $\Delta$ sector is decoupled 
from the off-diagonal variables, and so will have the same $\Delta^3\ln\Delta$ behavior as in Eq.~(\ref{gA0}). But the effort to obtain a full solution is so large that we abandoned further work on it.

\section{Hessian studies}
\label{sec:hessian}

In this section we write down the Hessian for the $m=3$ Heisenberg
spin glass in a form which is convenient for numerical work. The
Hessian is of interest as it describes the nature of the energy of the
spin glass in the vicinity of the minima. It is also closely related
to the matrices needed to describe the spin waves in the system
\cite{bm1981}.  We follow the approach used in the paper of Beton and
Moore~\cite{beton1984electron} to find the elements of the Hessian
matrix $T$ corresponding to directions transverse to each spin subject
to the above metastability condition. We first define the
site-dependent two-dimensional orthogonal unit vectors
$\hat{e}_{x}(i)$ and $\hat{e}_{y}(i)$ such that
\begin{align}
\hat{e}_{m}(i)\cdot\mathbf{S}_{i}^{0} &= 0\\
\hat{e}_{m}(i)\cdot\hat{e}_{n}(i) &= \delta^{mn},
\end{align}
where $m,n = x,y$ denotes the directions perpendicular to the spin at
the $i$th site, which is deemed in the $\lq \lq z"$ direction. The
linear combinations
$\hat{e}_{i}^{\pm}=\frac{1}{\sqrt{2}}(\hat{e}_{x}(i)\pm
i\hat{e}_{y}(i))$ turn out to be particularly useful. Expanding
$\mathbf{S}_{i}$ about $\mathbf{S}_{i}^{0}$, subject to the condition
that the length of the spins remains unchanged yields, upto
second-order:
\begin{align}
\mathbf{S}_{i} = \mathbf{S}_{i}^{0}+\Gamma_{i}^{x} \hat{e}_{x}(i)+\Gamma_{i}^{y} \hat{e}_{y}(i)-\frac{1}{2}[(\Gamma_{i}^{x})^{2}+(\Gamma_{i}^{y})^{2}]\mathbf{S}_{i}^{0}.
\end{align}
Equivalently, 
\begin{align}
\mathbf{S}_{i} = \mathbf{S}_{i}^{0}+Z_{i}^{-}\hat{e}_{i}^{+}+Z_{i}^{+}\hat{e}_{i}^{-}-Z_{i}^{-}Z_{i}^{+}\mathbf{S}_{i}^{0},
\end{align}
where $Z_{i}^{\pm} = \frac{1}{\sqrt{2}}(\Gamma_{i}^{x}\pm i\Gamma_{i}^{y})$, and $(Z_{i}^{+})^{*}=Z_{i}^{-}$. Defining the $2N$-dimensional vector 
\begin{align}
|Z\rangle = \begin{pmatrix} Z_{i}^{-}\\ Z_{i}^{+} \end{pmatrix},
\end{align}
the change in energy per spin component degree of freedom $\frac{\delta E}{3}$ due to a change in spin orientations $|Z\rangle$, is given by:
\begin{align}
\frac{\delta E}{3} = \frac{1}{2}\langle Z|T|Z\rangle,
\end{align}
where $T$ is the $2N \times 2N$ Hessian matrix given by
\begin{align*}
\begin{aligned}
T = \frac{1}{3}
\begin{pmatrix}
  |\mathbf{H}_{i}|\delta_{ij}+A_{ij}^{*}  & B_{ij}^{*}\\
  B_{ij}  & |\mathbf{H}_{i}|\delta_{ij}+A_{ij}
\end{pmatrix}
\end{aligned},
\end{align*}
where the matrix elements are 
\begin{align*}
A_{ij} = A_{ji}^{*} = -3J_{ij}\hat{e}_{i}^{+}\cdot\hat{e}_{j}^{-}\\
B_{ij} = B_{ji}^{*} = -3J_{ij}\hat{e}_{i}^{+}\cdot\hat{e}_{j}^{+}.
\end{align*}

Converting to spherical coordinates, the matrix elements are 
\begin{widetext}
\begin{align}
A_{ij}^{*} &= - \frac{3J_{ij}}{2}[(\cos(\theta_{i})\cos(\theta_{j})+1)\cos(\phi_{i}-\phi_{j})+i(\cos(\theta_{i})+\cos(\theta_{j}))\sin(\phi_{i}-\phi_{j})+\sin(\theta_{i})\sin(\theta_{j})]\nonumber\nonumber\\
B_{ij}^{*} &= - \frac{3J_{ij}}{2}[(\cos(\theta_{i})\cos(\theta_{j})-1)\cos(\phi_{i}-\phi_{j})-i(\cos(\theta_{i})-\cos(\theta_{j}))\sin(\phi_{i}-\phi_{j})+\sin(\theta_{i})\sin(\theta_{j})]\nonumber\\
B_{ij} &= - \frac{3J_{ij}}{2}[(\cos(\theta_{i})\cos(\theta_{j})-1)\cos(\phi_{i}-\phi_{j})+i(\cos(\theta_{i})-\cos(\theta_{j}))\sin(\phi_{i}-\phi_{j})+\sin(\theta_{i})\sin(\theta_{j})]\nonumber\\
A_{ij} &= - \frac{3J_{ij}}{2}[(\cos(\theta_{i})\cos(\theta_{j})+1)\cos(\phi_{i}-\phi_{j})-i(\cos(\theta_{i})+\cos(\theta_{j}))\sin(\phi_{i}-\phi_{j})+\sin(\theta_{i})\sin(\theta_{j})]\nonumber\\
\end{align}.
\end{widetext}

\begin{figure}
  \includegraphics[width=\columnwidth]{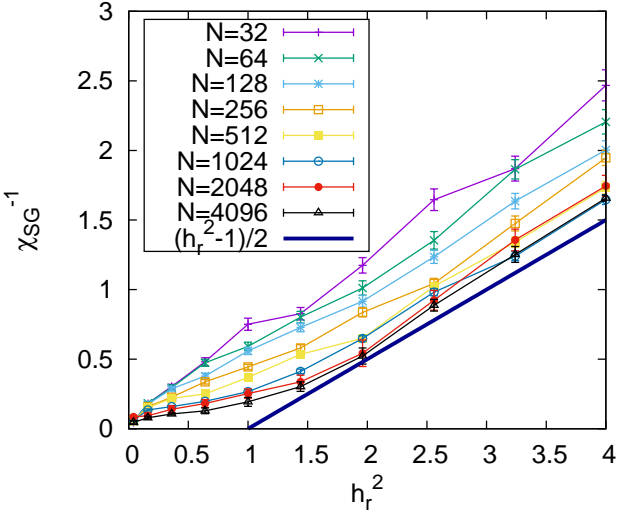} 
  \caption{(Color online) The inverse of the spin glass susceptibility $\chi_{SG}^{-1}$ versus $h_r^2$ for a range of system sizes of the Heisenberg SK model. The analytic curve is the result of Eq. (\ref{chisgexact}). For $h_r \le 1$, one expects that $\chi_{SG}^{-1} =0$, but finite size effects make it non-zero.}
  \label{fig1SK}
\end{figure}

\begin{figure}
  \includegraphics[width=\columnwidth]{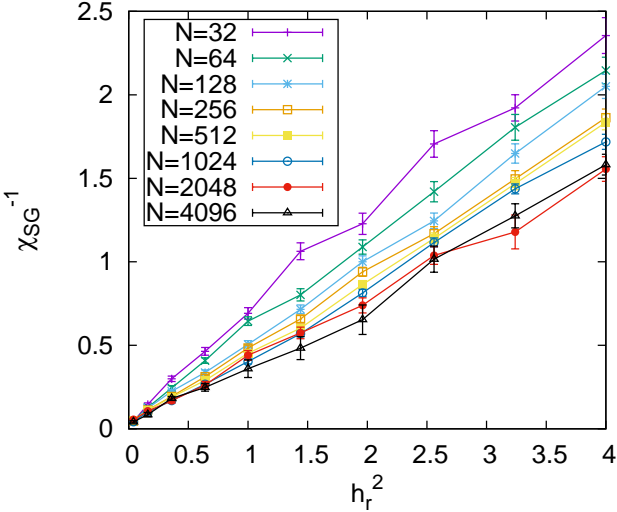} 
  \caption{(Color online) The inverse of the spin glass susceptibility $\chi_{SG}^{-1}$ versus $h_r^2$ for a range of system sizes for the VB model with $z = 6$.}
  \label{fig1VB}
\end{figure}

In the next subsection we use the Hessian to numerically calculate the spin glass susceptibility of both the SK model and VB model in a range of random fields for the Heisenberg spin glass.

\subsection{Spin Glass Susceptibility}
\label{sec:spinglasssusceptibility}
The spin glass susceptibility for the metastable states can be
computed from the inverse of the Hessian matrix using the
relation~\cite{bm1981}
\begin{equation}
  \chi_{SG} = \frac{1}{N}\Tr{(T^{-1})^{2}}.
\end{equation}
For the SK model and $h_r> h_{AT}=1$, we have calculated $\chi_{SG}$
analytically and Fig. \ref{fig1SK} shows that our numerical work is
approaching the analytical solution, but finite size effects are still
very considerable at the sizes we can study. Notice that for the SK
model there is (weak) numerical evidence that $\chi_{SG}$ diverges
below the AT field. For the VB model, the plot of $\chi_{SG}$ in
Fig. \ref{fig1VB} obtained from our metastable states which lie above
the true ground state energy provides no evidence that an AT field has
much relevance for these states.

\begin{figure}
  \includegraphics[width=\columnwidth]{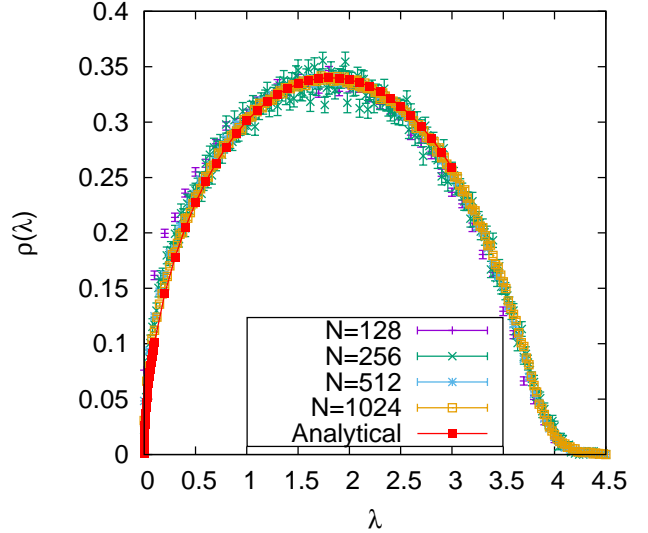} 
  \caption{(Color online) The averaged density of states of the Hessian matrix of the metastable states obtained after a quench to $T=0$ starting from spins with random orientations i.e. $T= \infty$ for the SK model ($\sigma =0, z =N-1$ of the diluted model). Data shown here for the special case of $h_{r} = 0.8$, for which the system is in the spin glass phase, just below $h_{AT} = 1$. The analytical curve is that calculated from Eqs. (\ref{eqn:rho}) and (\ref{Gbar})) for metastable states at the top of the band within the annealed approximation. The numerical results are strikingly similar to the analytical results, despite the fact that they refer to Hessians for quite different situations! }
  \label{fig2}
\end{figure}

\subsection{Density of States}
\label{sec:density}

The density of states of the eigenvalues of the  Hessian matrix has been obtained numerically for
the minima obtained in a quench from infinite temperature to zero
temperature. The results have remarkable agreement with the analytical
calculation performed on the Heisenberg SK model as shown in Fig.~\ref{fig2}. The
analytical calculation itself is not for the same set of metastable
states.  It applies to the states corresponding to $u=0$ (i.e. those
with the largest complexity within the annealed approximation).  In
Fig.~\ref{fig2}, data are shown for $h_r =0.8 h_{AT}$, where no gap is
present.  The agreement between the analytical curve which is obtained
for the thermodynamic limit, and the data for a $N=1024$ size system
from numerical simulations, is striking. Notice that the $\sqrt{\lambda}$ form predicted from the annealed study (see Sec. \ref{SKannealed}) seems to hold as $\lambda \to 0$, despite there being no Goldstone theorem in the presence of a random field to ensure the existence of massless modes.

We have also studied the density of states and quantities like the
inverse participation ratios for the quenched state minima in models
like the VB model and the one-dimensional long range models. Basically
the results seem similar to those reported in
Refs. \onlinecite{baity2015soft} for the three dimensional Heisenberg
spin glass model in a random field. But it requires large systems to
get accurate results for the density of states at small values of
$\lambda$ and we are leaving these issues to a future publication.

\section{Conclusions}
\label{sec:conclusions}
We believe that the most interesting feature which has turned up in
our studies is the discovery for the SK model in zero external fields
that the quenched states reached for $m=2$ and $m =3$ are quite close
to the critical energies $E_c$ at which the overlap of the states
would acquire features associated with a $P(q)$ with broken replica
symmetry. In the Ising SK model the local fields after the quench are
so different from those used in the analytical calculations of $E_c$
that the connection of the quenched state to being just at the edge of
the states with broken replica symmetry was not recognized. Thus in
systems with marginal stability this means that features normally
associated with continuous phase transitions, in particular diverging
length scales, could be studied as in Ref. \onlinecite{bm:81a}. 

We have noticed too that the energy of the states reached from the quench
have zero overlap with each other. This behavior was predicted for the Ising case in Ref.~\onlinecite{newman:99} by  Newman and Stein who proved that after a
quench from infinite temperature for Ising systems the states which
are reached have a characteristic energy and a trivial $P(q)$. It
would be good to extend their theorems to vector spin systems both in
zero field and also in the presence of random fields.

In Sec. \ref{sec:SKanalytic} we attempted to extend the old calculations of Bray and Moore \cite{bm1981} which were for zero random field to non-zero random fields. For fields $h_r >h_{AT}$ where the complexity is zero, the annealed approximation is exact and we were able to obtain the exact form for the behaviour of the density of states of the Hessian matrix. There was found to be a gap in the spectrum which went to zero in the limit $h_r \to h_{AT}$. When $h_r< h_{AT}$ one needs to study the quenched average in order to get results pertinent to typical minima, but we were not able to overcome the algebraic complexities (see Sec. \ref{sec:quenched} and the Appendix), although the only difficulty is that of  solving the equations which we have obtained. If that could be done then one could investigate the limit of stability of the replica symmetric solution and determine
$E_c(h_r)$. Then one could investigate whether a quench in a field $h_r$ takes one to the limit of stability towards full replica symmetry breaking i.e.  $E_c(h_r)$, just as we found for $h_r=0$.

The annealed approximation is  tractable but alas it is only an approximation. Nevertheless the studies in Sec. \ref{sec:density} shows that it gives good results for the density of states of the Hessian for the SK model for $h_r < h_{AT}$. 

The VB model is a mean-field model and one could hope that it too could be understood analytically, but we do not know how this might be achieved. Our numerical studies of the density of states of its Hessian indicates that this is very different from that of the SK model. This is probably because for the SK model all the eigenstates are extended, whereas for the VB model, eigenvectors can also be localized. In fact our results for the VB model are quite similar to those reported for the three dimensional Heisenberg spin glass in a field \cite{baity2015soft}. There seems to be localized states lying in the gap region, all the way down to $\lambda=0$.  But understanding the VB model analytically is very challenging.

\acknowledgements
We should like to thank the authors of Ref.~\onlinecite{lupo:16} for an advance copy of their paper and  helpful discussions. One of us (MAM) would like to thank Dan Stein for discussions on quenches in  Ising systems. AS acknowledges support from the DST-INSPIRE Faculty Award [DST/INSPIRE/04/2014/002461]. JY was supported by
Basic Science Research Program through the National Research Foundation 
of Korea (NRF) funded by the Ministry of Education (2014R1A1A2053362).

\begin{widetext}
\appendix*

\section{The quenched complexity details}
We present in this Appendix some of the details of the calculation of the quenched complexity $g(\varepsilon)=N^{-1}\langle\ln N_S(\varepsilon)\rangle$.
We first replicate Eq.~(\ref{NS}) to obtain 
\begin{align}
 [N_S(\varepsilon)]^n=&\int\prod_{i,\alpha,\eta}dH^\eta_{i\alpha}
 \int\prod_{i,\alpha,\eta}\frac{dx^\eta_{i\alpha}}{2\pi i}\int \prod_{\eta}\frac{du^\eta}{2\pi i}\;
 \exp\left[ \sum_{i,\alpha,\eta}x^\eta_{i\alpha}H^\eta_{i\alpha}-\sum_{i<j,\alpha,\eta}J_{ij}
 (x^\eta_{i\alpha}\hat{H}^\eta_{j\alpha}
+x^\eta_{j\alpha}\hat{H}^\eta_{i\alpha})-\sum_{i,\alpha,\eta}x^\eta_{i\alpha}h^{\rm ex}_{i\alpha}\right]
\nonumber \\
&~~~~~\times \prod_\eta |\det\mathsf{M}^\eta \{J_{ij}\}|
\exp\left[ -\sum_\eta u^\eta Nm\varepsilon
-\sum_\eta \frac 1 2 u^\eta  m\sum_i (H^\eta_i+ \hat{\bm{H}}^\eta_i\cdot\bm{h}^{\rm ex}_i )
\right],
\end{align}
where $i,j,\ldots=1,\ldots,N$ are the site indices, $\alpha,\beta,\ldots=1,\ldots,m$ the vector component indices, and
$\eta,\mu,\nu,\ldots=1,\ldots,n$ replica indices.
The average over $J_{ij}$ can be done as in Ref.~\onlinecite{bm1981}. We have
\begin{align}
& \left\langle 
 \exp\left[-\sum_{i<j}J_{ij}\sum_{i,\alpha} (x^\eta_{i\alpha}\hat{H}^\eta_{j\alpha}
+x^\eta_{j\alpha}\hat{H}^\eta_{i\alpha})\right]  \prod_\eta |\det\mathsf{M}^\eta \{J_{ij}\}| 
 \right\rangle_J \nonumber \\
=&\exp \left[ \frac 1 {2N} \sum_{i<j}\left\{ \sum_{\alpha,\eta}(x^\eta_{i\alpha}\hat{H}^\eta_{j\alpha}
+x^\eta_{j\alpha}\hat{H}^\eta_{i\alpha}) \right\}^2\right] 
\left\langle \prod_\eta |\det\mathsf{M}^\eta \{J_{ij}-O(\frac 1 N)\}|  \right\rangle_J
 \end{align}
After neglecting the $O(1/N)$ term, the determinant can be evaluated to yield the replicated version of 
Eq.~(\ref{det}). Using the Hubbard-Stratonovich transformation
and the rotational invariance, we can write the exponential factor in front of the determinant as
\begin{align}  
 &\exp\left[\frac 1{2m}\sum_{i,\alpha,\eta}(x^\eta_{i\alpha})^2 \right]
 \int\prod_\eta\frac{dv^\eta}{(2\pi/Nm)^{1/2}}\;
 \exp\left[ -\frac{Nm}{2}\sum_\eta (v^\eta)^2+\sum_\eta v^\eta 
 \left( \sum_{i,\alpha}x^\eta_{i\alpha}\hat{H}^\eta_{i\alpha}\right) \right]
 \nonumber \\
\times &\int\prod_{\eta<\nu}\frac{dA_{\eta\nu}dA^*_{\eta\nu}}{(\pi/Nm)}\; 
\exp\left[ -Nm \sum_{\eta<\nu}|A_{\eta\nu}|^2 
+\sum_{\eta<\nu}A^*_{\eta\nu}\left( \sum_{i,\alpha}x^\eta_{i\alpha}x^\nu_{i\alpha}\right)
+\sum_{\eta<\nu}A_{\eta\nu} \left( \sum_{i,\alpha}\hat{H}^\eta_{i\alpha}\hat{H}^\nu_{i\alpha}\right)
\right]
\nonumber \\
\times &\int\prod_{\eta<\nu}\frac{dB_{\eta\nu}dB^*_{\eta\nu}}{(\pi/Nm)}\; 
\exp\left[ -Nm \sum_{\eta<\nu}|B_{\eta\nu}|^2 
+\sum_{\eta<\nu}B^*_{\eta\nu}\left( \sum_{i,\alpha}x^\eta_{i\alpha}\hat{H}^\nu_{i\alpha}\right)
+\sum_{\eta<\nu}B_{\eta\nu} \left( \sum_{i,\alpha}\hat{H}^\eta_{i\alpha}x^\nu_{i\alpha}\right)
\right]
\end{align}
The average over the random external field is done as
\begin{align}
& \left\langle 
 \exp\left[-\sum_{i,\alpha,\eta} (x^\eta_{i\alpha}+\frac 1 2 u^\eta m \hat{H}^\eta_{i\alpha})h^{\rm ex}_{i\alpha}\right]
 \right\rangle_{\bm{h}^{\rm ex}}  \\
=&\exp\left[ 
\frac{h^2_r}{2m}\sum_{i,\alpha,\eta}(x^\eta_{i\alpha})^2
+\frac{h^2_r}{2}\sum_\eta u^\eta  \left( \sum_{i,\alpha}x^\eta_{i\alpha}\hat{H}^\eta_{i\alpha}\right)
+Nm\frac{h^2_r }{8}\sum_\eta (u^\eta)^2
\right] \nonumber  \\
\times&
\exp\left[
\frac{h^2_r}{m}\sum_{\eta<\nu}\left\{
\left( \sum_{i,\alpha}x^\eta_{i\alpha}x^\nu_{i\alpha}\right)
+\frac{m}{2}u^\nu \left( \sum_{i,\alpha}x^\eta_{i\alpha}\hat{H}^\nu_{i\alpha}\right)
+\frac{m}{2}u^\eta  \left( \sum_{i,\alpha}\hat{H}^\eta_{i\alpha}x^\nu_{i\alpha}\right)
+\frac{m^2}{4}u^\eta u^\nu  \left( \sum_{i,\alpha}\hat{H}^\eta_{i\alpha}\hat{H}^\nu_{i\alpha}\right)
\right\}
\right].
\nonumber 
 \end{align}

All the site indices are now decoupled. Using the delta function constraint for $\bar{\chi}$, we have
\begin{align}
 \langle [N_S(\varepsilon)]^n \rangle_{J,h^{\rm ex}}=&
 \int\prod_\eta du^\eta \int\prod_\eta dv^\eta\int\prod_\eta d\bar{\chi}^\eta\int\prod_\eta d\lambda^\eta
 \int\prod_{\eta<\nu}dA_{\eta\nu}dA^*_{\eta\nu}
 \int\prod_{\eta<\nu}dB_{\eta\nu}dB^*_{\eta\nu}
  \nonumber \\
&\times 
\exp\Bigg[
Nm\sum_\eta \lambda^\eta \bar{\chi}^\eta + \frac{Nm}{2}\sum_\eta (\bar{\chi}^\eta)^2 
-Nm\varepsilon \sum_\eta u^\eta -\frac{Nm}{2}\sum_\eta (v^\eta)^2
+Nm\frac{h^2_r}{8}\sum_\eta (u^\eta)^2 \nonumber \\
&~~~~~~~~-Nm \sum_{\eta<\nu}(|A_{\eta\nu}|^2+|B_{\eta\nu}|^2)+N\ln I 
\Bigg], 
 \end{align}
where
\begin{align}
 I=&\int\prod_{\eta,\alpha}dH_\alpha^\eta
 \int\prod_{\eta,\alpha}dx_\alpha^\eta 
 \left(1-\frac{\bar{\chi}^\eta}{H^\eta}\right)^{m-1}  \nonumber \\
 &\times \exp\Bigg[
 \frac{1+h^2_r}{2m}\sum_{\eta\alpha}(x_\alpha^\eta)^2
+ \sum_\eta (v^\eta+\frac{h^2_r}{2}u^\eta)\sum_\alpha x^\eta_\alpha\hat{H}^\eta_\alpha 
 +\sum_{\eta,\alpha}x^\eta_\alpha H^\eta_\alpha -(m-1)\sum_\eta \lambda^\eta (H^\eta-\bar{\chi}^\eta)^{-1}
 -\frac{m}{2}\sum_\eta u^\eta H^\eta 
 \Bigg] \nonumber \\
 &\times
 \exp\Bigg[ 
 \sum_{\eta<\nu}(A^*_{\eta\nu}+\frac{h^2_r}{m}) \left(\sum_\alpha x^\eta_\alpha x^\nu_\alpha\right) 
 +\sum_{\eta<\nu}(A_{\eta\nu}+\frac{mh^2_r}{4}u^\eta u^\nu )
 \left(\sum_\alpha \hat{H}^\eta_\alpha \hat{H}^\nu_\alpha \right)\Bigg] 
 \nonumber \\
 &\times
 \exp\Bigg[ 
 \sum_{\eta<\nu}(B^*_{\eta\nu}+\frac{h^2_r}{2}u^\nu) 
 \left(\sum_\alpha x^\eta_\alpha \hat{H}^\nu_\alpha\right) 
 +\sum_{\eta<\nu}(B_{\eta\nu}+\frac{h^2_r}{2}u^\eta )
 \left(\sum_\alpha \hat{H}^\eta_\alpha x^\nu_\alpha \right)
 \Bigg]
\end{align}
The saddle point equation for the off-diagonal variables are given by 
\begin{equation}
 A_{\eta\nu}=\frac{1}{m}\langle\sum_\alpha x^\eta_\alpha x^\nu_\alpha \rangle ,~~
  A^*_{\eta\nu}=\frac{1}{m}\langle\sum_\alpha \hat{H}^\eta_\alpha \hat{H}^\nu_\alpha \rangle,~~
   B_{\eta\nu}=\frac{1}{m}\langle\sum_\alpha x^\eta_\alpha \hat{H}^\nu_\alpha \rangle,~~
    B^*_{\eta\nu}=\frac{1}{m}\langle\sum_\alpha \hat{H}^\eta_\alpha x^\nu_\alpha \rangle,
    \label{saddle}
\end{equation}
where $\langle~\rangle$ is calculated with respect to $I$.
We can easily see that when $h_r\neq 0$, these averages do not become zero even when
all the integration variables carrying off-diagonal replica indices vanish.  
Therefore $A_{\eta\nu}=A^*_{\eta\nu}=B_{\eta\nu}=B^*_{\eta\nu}=0$
is not a solution of the saddle point equations.

We now study the saddle points in the replica symmetric form,
\begin{equation}
A_{\eta\nu}=A,~~A^*_{\eta\nu}=A^*,~~B_{\eta\nu}=B^*_{\eta\nu}=B,~~u^\eta=u,~~v^\eta=v,~~
\bar{\chi}^\eta=\bar{\chi},~~\lambda^\eta=\lambda.
\end{equation}
Then 
\begin{equation}
g(\varepsilon)=N^{-1}\langle \ln N_S(\varepsilon) \rangle_{J,h^{\rm ex}}=
m \Big\{ \lambda \bar{\chi} + \frac{1}{2} \bar{\chi}^2 
-\varepsilon  u -\frac{1}{2}v^2 
+\frac{h^2_r}{8} u^2 +\frac{1}{2} (|A|^2+B^2)\Big\} +\lim_{n\to 0}[\frac{1}{n}\ln I ],
\end{equation}
where 
\begin{align}
 I=&\int\prod_{\eta,\alpha}dH_\alpha^\eta
 \int\prod_{\eta,\alpha}dx_\alpha^\eta 
 \left(1-\frac{\bar{\chi}}{H^\eta}\right)^{m-1}  \\
 &\times \exp\Bigg[
 \frac{1+h^2_r}{2m}\sum_{\eta,\alpha}(x_\alpha^\eta)^2
+ (v+\frac{h^2_r}{2}u)\sum_{\eta,\alpha} x^\eta_\alpha\hat{H}^\eta_\alpha 
 +\sum_{\eta,\alpha}x^\eta_\alpha H^\eta_\alpha -(m-1)\lambda\sum_\eta  (H^\eta-\bar{\chi})^{-1}
 -\frac{m}{2}u\sum_\eta H^\eta 
 \Bigg] \nonumber \\
 &\times
 \exp\Bigg[ 
 (A^*+\frac{h^2_r}{m}) \sum_{\eta<\nu}\sum_\alpha x^\eta_\alpha x^\nu_\alpha
 +(A+\frac{mh^2_r}{4}u^2 )\sum_{\eta<\nu}
 \sum_\alpha \hat{H}^\eta_\alpha \hat{H}^\nu_\alpha \Bigg] 
 \nonumber \\
 &\times
 \exp\Bigg[ 
 (B+\frac{h^2_r}{2}u) 
\sum_{\eta<\nu} \left(\sum_\alpha x^\eta_\alpha \hat{H}^\nu_\alpha
+\sum_\alpha \hat{H}^\eta_\alpha x^\nu_\alpha \right)
 \Bigg].
 \nonumber 
\end{align}

We now use the Hubbard-Stratonovich transformations on the last three terms 
in the previous equation
using the auxiliary variables, $w_\alpha$, $y_\alpha$, $z_\alpha$
and $z^*_\alpha$, to disentangle the replica indices. Then we can write 
\begin{equation}
 \int\prod_{\eta,\alpha}dH_\alpha^\eta
 \int\prod_{\eta,\alpha}dx_\alpha^\eta \; \sum_\eta(\cdots) =
 \left[ \int\prod_{\alpha}dH_\alpha
 \int\prod_{\alpha}dx_\alpha (\cdots) \right]^n.
\end{equation}
By explicitly evaluating $\lim_{n\to 0}n^{-1}\ln I$, we obtain 
\begin{align}
g(\varepsilon)=&
m \Big\{ \lambda \bar{\chi} + \frac{1}{2} \bar{\chi}^2 
-\varepsilon  u -\frac{1}{2}v^2 - \frac{A}{2m}
+\frac{1}{2} (AA^*+B^2)\Big\} \nonumber  \\
&+
\int\prod_\alpha \frac{dw_\alpha}{\sqrt{2\pi}}
\int\prod_\alpha \frac{dy_\alpha}{\sqrt{2\pi}}
\int\prod_\alpha \frac{dz_\alpha dz^*_\alpha}{2\pi}\; 
\exp[-\frac 1 2 \sum_\alpha (w^2_\alpha+y^2_\alpha+|z_\alpha|^2)]\; \ln J,
\end{align}
where
\begin{align}
 J=&
 \int\prod_{\alpha}dH_\alpha
 \int^{i\infty}_{-i\infty} \prod_{\alpha}\frac{dx_\alpha}{2\pi i} 
 \left(1-\frac{\bar{\chi}}{H}\right)^{m-1}  \\
 &\times \exp\Bigg[
 \frac{1-mA^*}{2m}\sum_{\alpha}(x_\alpha)^2
+ (v-B)\sum_{\alpha} x_\alpha\hat{H}_\alpha 
 +\sum_{\alpha}x_\alpha H_\alpha -(m-1)\lambda  (H-\bar{\chi})^{-1}
 -\frac{m}{2}uH 
 \Bigg] \nonumber \\
 &\times
 \exp\Bigg[ 
 \sqrt{A^*+\frac{h^2_r}{m}} \sum_\alpha w_\alpha x_\alpha
 +\sqrt{A+\frac{mh^2_r}{4}u^2 } \sum_\alpha y_\alpha  \hat{H}_\alpha 
+ \sqrt{\frac{1}{2}(B+\frac{h^2_r}{2}u) }
\sum_\alpha  \left(z_\alpha x_\alpha +z^*_\alpha \hat{H}_\alpha\right)
 \Bigg].
 \nonumber 
\end{align}

Now we change the integration variable in $J$ from $\bm{H}$ to 
$\bm{h}\equiv \bm{H}-\bar{\chi}\hat{\bm{H}}=(H-\bar{\chi})\hat{\bm{H}}$.
The lower limit of the integral for $\bm{h}$ now becomes 0 and the Jacobian exactly cancels the factor of
$(1-\bar{\chi}/H)^{m-1}$.  Let us also use the new variable $\Delta$, where $v=-\chi-\Delta$ so that $H+v=h-\Delta$. 
Extremizing with respect to $\chi$ in $g(\varepsilon)$ yields
$ \lambda-\Delta-u / 2 =0$.
We finally have  
\begin{align}
g(\varepsilon)=&
m \Big\{ -\frac{\Delta^2}{2} -\varepsilon u   - \frac{A}{2m}
+\frac{1}{2} (AA^*+B^2)\Big\} \nonumber  \\
&+
\int \frac{d^m \bm{w}}{(2\pi)^{m/2}}
\int \frac{d^m \bm{y}}{(2\pi)^{m/2}}
\int \frac{d^m \bm{z} d^m\bm{z}^*}{(2\pi)^m}\; 
\exp[-\frac 1 2 \sum_\alpha (w^2_\alpha+y^2_\alpha+|z_\alpha|^2)]\; \ln K(\bm{w},\bm{y},\bm{z},\bm{z}^*),
\label{ge}
\end{align}
where
\begin{align}
 K=&
 \int d^m\bm{h}
 \int^{i\infty}_{-i\infty} \frac{d^m \bm{x}}{2\pi i} 
 \;\exp\Bigg[
 \frac{1-mA^*}{2m}\bm{x}^2
+ (h-\Delta-B)\bm{x}\cdot\hat{\bm{h}}
  -(m-1)\frac{\Delta + u/2}{h}
 -\frac{m}{2}uh 
 \Bigg] \nonumber \\
 &\times
 \exp\Bigg[ 
 \sqrt{A^*+\frac{h^2_r}{m}} \;\bm{w}\cdot\bm{x}
 +\sqrt{A+\frac{mh^2_r}{4}u^2 }  \; \bm{y}\cdot  \hat{\bm{h}} 
+ \sqrt{\frac{1}{2}(B+\frac{h^2_r}{2}u) }
\left(\bm{z}\cdot\bm{x}+\bm{z}^* \cdot \hat{\bm{h}}\right)
 \Bigg].
\end{align}

\end{widetext}

\bibliography{refs}

\begin{thebibliography}{29}%
\makeatletter
\providecommand \@ifxundefined [1]{%
 \@ifx{#1\undefined}
}%
\providecommand \@ifnum [1]{%
 \ifnum #1\expandafter \@firstoftwo
 \else \expandafter \@secondoftwo
 \fi
}%
\providecommand \@ifx [1]{%
 \ifx #1\expandafter \@firstoftwo
 \else \expandafter \@secondoftwo
 \fi
}%
\providecommand \natexlab [1]{#1}%
\providecommand \enquote  [1]{``#1''}%
\providecommand \bibnamefont  [1]{#1}%
\providecommand \bibfnamefont [1]{#1}%
\providecommand \citenamefont [1]{#1}%
\providecommand \href@noop [0]{\@secondoftwo}%
\providecommand \href [0]{\begingroup \@sanitize@url \@href}%
\providecommand \@href[1]{\@@startlink{#1}\@@href}%
\providecommand \@@href[1]{\endgroup#1\@@endlink}%
\providecommand \@sanitize@url [0]{\catcode `\\12\catcode `\$12\catcode
  `\&12\catcode `\#12\catcode `\^12\catcode `\_12\catcode `\%12\relax}%
\providecommand \@@startlink[1]{}%
\providecommand \@@endlink[0]{}%
\providecommand \url  [0]{\begingroup\@sanitize@url \@url }%
\providecommand \@url [1]{\endgroup\@href {#1}{\urlprefix }}%
\providecommand \urlprefix  [0]{URL }%
\providecommand \Eprint [0]{\href }%
\providecommand \doibase [0]{http://dx.doi.org/}%
\providecommand \selectlanguage [0]{\@gobble}%
\providecommand \bibinfo  [0]{\@secondoftwo}%
\providecommand \bibfield  [0]{\@secondoftwo}%
\providecommand \translation [1]{[#1]}%
\providecommand \BibitemOpen [0]{}%
\providecommand \bibitemStop [0]{}%
\providecommand \bibitemNoStop [0]{.\EOS\space}%
\providecommand \EOS [0]{\spacefactor3000\relax}%
\providecommand \BibitemShut  [1]{\csname bibitem#1\endcsname}%
\let\auto@bib@innerbib\@empty
\bibitem [{\citenamefont {{Charbonneau}}\ \emph {et~al.}(2016)\citenamefont
  {{Charbonneau}}, \citenamefont {{Kurchan}}, \citenamefont {{Parisi}},
  \citenamefont {{Urbani}},\ and\ \citenamefont {{Zamponi}}}]{charbonneau16}%
  \BibitemOpen
  \bibfield  {author} {\bibinfo {author} {\bibfnamefont {P.}~\bibnamefont
  {{Charbonneau}}}, \bibinfo {author} {\bibfnamefont {J.}~\bibnamefont
  {{Kurchan}}}, \bibinfo {author} {\bibfnamefont {G.}~\bibnamefont {{Parisi}}},
  \bibinfo {author} {\bibfnamefont {P.}~\bibnamefont {{Urbani}}}, \ and\
  \bibinfo {author} {\bibfnamefont {F.}~\bibnamefont {{Zamponi}}},\ }\href@noop
  {} {\bibfield  {journal} {\bibinfo  {journal} {eprint
  arXiv:cond-mat/1605.03008}\ } (\bibinfo {year} {2016})}\BibitemShut {NoStop}%
\bibitem [{\citenamefont {{Baule}}\ \emph {et~al.}(2016)\citenamefont
  {{Baule}}, \citenamefont {{Morone}}, \citenamefont {{Herrmann}},\ and\
  \citenamefont {{Makse}}}]{baule16}%
  \BibitemOpen
  \bibfield  {author} {\bibinfo {author} {\bibfnamefont {A.}~\bibnamefont
  {{Baule}}}, \bibinfo {author} {\bibfnamefont {F.}~\bibnamefont {{Morone}}},
  \bibinfo {author} {\bibfnamefont {H.~J.}\ \bibnamefont {{Herrmann}}}, \ and\
  \bibinfo {author} {\bibfnamefont {H.~A.}\ \bibnamefont {{Makse}}},\
  }\href@noop {} {\bibfield  {journal} {\bibinfo  {journal} {e-print
  arXiv:cond-mat/1602.04369}\ } (\bibinfo {year} {2016})}\BibitemShut {NoStop}%
\bibitem [{\citenamefont {Tarzia}\ and\ \citenamefont
  {Moore}(2007)}]{tarzia2007glass}%
  \BibitemOpen
  \bibfield  {author} {\bibinfo {author} {\bibfnamefont {M.}~\bibnamefont
  {Tarzia}}\ and\ \bibinfo {author} {\bibfnamefont {M.~A.}\ \bibnamefont
  {Moore}},\ }\href {\doibase 10.1103/PhysRevE.75.031502} {\bibfield  {journal}
  {\bibinfo  {journal} {Phys. Rev. E}\ }\textbf {\bibinfo {volume} {75}},\
  \bibinfo {pages} {031502} (\bibinfo {year} {2007})}\BibitemShut {NoStop}%
\bibitem [{\citenamefont {Fullerton}\ and\ \citenamefont
  {Moore}(2013)}]{fullerton2013growing}%
  \BibitemOpen
  \bibfield  {author} {\bibinfo {author} {\bibfnamefont {C.~J.}\ \bibnamefont
  {Fullerton}}\ and\ \bibinfo {author} {\bibfnamefont {M.~A.}\ \bibnamefont
  {Moore}},\ }\href@noop {} {\bibfield  {journal} {\bibinfo  {journal} {eprint
  arXiv:cond-mat/1304.4420}\ } (\bibinfo {year} {2013})}\BibitemShut {NoStop}%
\bibitem [{\citenamefont {Moore}\ and\ \citenamefont {Yeo}(2006)}]{moore06}%
  \BibitemOpen
  \bibfield  {author} {\bibinfo {author} {\bibfnamefont {M.~A.}\ \bibnamefont
  {Moore}}\ and\ \bibinfo {author} {\bibfnamefont {J.}~\bibnamefont {Yeo}},\
  }\href {\doibase 10.1103/PhysRevLett.96.095701} {\bibfield  {journal}
  {\bibinfo  {journal} {Phys. Rev. Lett.}\ }\textbf {\bibinfo {volume} {96}},\
  \bibinfo {pages} {095701} (\bibinfo {year} {2006})}\BibitemShut {NoStop}%
\bibitem [{\citenamefont {Baity-Jesi}\ \emph {et~al.}(2015)\citenamefont
  {Baity-Jesi}, \citenamefont {Mart{\'\i}n-Mayor}, \citenamefont {Parisi},\
  and\ \citenamefont {Perez-Gaviro}}]{baity2015soft}%
  \BibitemOpen
  \bibfield  {author} {\bibinfo {author} {\bibfnamefont {M.}~\bibnamefont
  {Baity-Jesi}}, \bibinfo {author} {\bibfnamefont {V.}~\bibnamefont
  {Mart{\'\i}n-Mayor}}, \bibinfo {author} {\bibfnamefont {G.}~\bibnamefont
  {Parisi}}, \ and\ \bibinfo {author} {\bibfnamefont {S.}~\bibnamefont
  {Perez-Gaviro}},\ }\href@noop {} {\bibfield  {journal} {\bibinfo  {journal}
  {Physical Review Letters}\ }\textbf {\bibinfo {volume} {115}},\ \bibinfo
  {pages} {267205} (\bibinfo {year} {2015})}\BibitemShut {NoStop}%
\bibitem [{\citenamefont {Wyart}\ \emph {et~al.}(2005)\citenamefont {Wyart},
  \citenamefont {Nagel},\ and\ \citenamefont {Witten}}]{wyart2005geometric}%
  \BibitemOpen
  \bibfield  {author} {\bibinfo {author} {\bibfnamefont {M.}~\bibnamefont
  {Wyart}}, \bibinfo {author} {\bibfnamefont {S.~R.}\ \bibnamefont {Nagel}}, \
  and\ \bibinfo {author} {\bibfnamefont {T.~A.}\ \bibnamefont {Witten}},\
  }\href@noop {} {\bibfield  {journal} {\bibinfo  {journal} {EPL (Europhysics
  Letters)}\ }\textbf {\bibinfo {volume} {72}},\ \bibinfo {pages} {486}
  (\bibinfo {year} {2005})}\BibitemShut {NoStop}%
\bibitem [{\citenamefont {{Charbonneau}}\ \emph {et~al.}(2015)\citenamefont
  {{Charbonneau}}, \citenamefont {{Corwin}}, \citenamefont {{Parisi}},
  \citenamefont {{Poncet}},\ and\ \citenamefont {{Zamponi}}}]{charbonneau15}%
  \BibitemOpen
  \bibfield  {author} {\bibinfo {author} {\bibfnamefont {P.}~\bibnamefont
  {{Charbonneau}}}, \bibinfo {author} {\bibfnamefont {E.~I.}\ \bibnamefont
  {{Corwin}}}, \bibinfo {author} {\bibfnamefont {G.}~\bibnamefont {{Parisi}}},
  \bibinfo {author} {\bibfnamefont {A.}~\bibnamefont {{Poncet}}}, \ and\
  \bibinfo {author} {\bibfnamefont {F.}~\bibnamefont {{Zamponi}}},\ }\href@noop
  {} {\bibfield  {journal} {\bibinfo  {journal} {ArXiv e-prints}\ } (\bibinfo
  {year} {2015})},\ \Eprint {http://arxiv.org/abs/1512.09100} {arXiv:1512.09100
  [cond-mat.dis-nn]} \BibitemShut {NoStop}%
\bibitem [{\citenamefont {Bray}\ and\ \citenamefont
  {Moore}(1981{\natexlab{a}})}]{bm1981}%
  \BibitemOpen
  \bibfield  {author} {\bibinfo {author} {\bibfnamefont {A.~J.}\ \bibnamefont
  {Bray}}\ and\ \bibinfo {author} {\bibfnamefont {M.~A.}\ \bibnamefont
  {Moore}},\ }\href@noop {} {\bibfield  {journal} {\bibinfo  {journal} {Journal
  of Physics C: Solid State Physics}\ }\textbf {\bibinfo {volume} {14}},\
  \bibinfo {pages} {2629} (\bibinfo {year} {1981}{\natexlab{a}})}\BibitemShut
  {NoStop}%
\bibitem [{\citenamefont {Yeo}\ and\ \citenamefont {Moore}(2004)}]{yeo04}%
  \BibitemOpen
  \bibfield  {author} {\bibinfo {author} {\bibfnamefont {J.}~\bibnamefont
  {Yeo}}\ and\ \bibinfo {author} {\bibfnamefont {M.~A.}\ \bibnamefont
  {Moore}},\ }\href {\doibase 10.1103/PhysRevLett.93.077201} {\bibfield
  {journal} {\bibinfo  {journal} {Phys. Rev. Lett.}\ }\textbf {\bibinfo
  {volume} {93}},\ \bibinfo {pages} {077201} (\bibinfo {year}
  {2004})}\BibitemShut {NoStop}%
\bibitem [{\citenamefont {Bray}\ and\ \citenamefont {Moore}(1982)}]{bm1982}%
  \BibitemOpen
  \bibfield  {author} {\bibinfo {author} {\bibfnamefont {A.~J.}\ \bibnamefont
  {Bray}}\ and\ \bibinfo {author} {\bibfnamefont {M.~A.}\ \bibnamefont
  {Moore}},\ }\href {http://stacks.iop.org/0022-3719/15/i=11/a=021} {\bibfield
  {journal} {\bibinfo  {journal} {Journal of Physics C: Solid State Physics}\
  }\textbf {\bibinfo {volume} {15}},\ \bibinfo {pages} {2417} (\bibinfo {year}
  {1982})}\BibitemShut {NoStop}%
\bibitem [{\citenamefont {Leuzzi}\ \emph {et~al.}(2008)\citenamefont {Leuzzi},
  \citenamefont {Parisi}, \citenamefont {Ricci-Tersenghi},\ and\ \citenamefont
  {Ruiz-Lorenzo}}]{leuzzi2008dilute}%
  \BibitemOpen
  \bibfield  {author} {\bibinfo {author} {\bibfnamefont {L.}~\bibnamefont
  {Leuzzi}}, \bibinfo {author} {\bibfnamefont {G.}~\bibnamefont {Parisi}},
  \bibinfo {author} {\bibfnamefont {F.}~\bibnamefont {Ricci-Tersenghi}}, \ and\
  \bibinfo {author} {\bibfnamefont {J.}~\bibnamefont {Ruiz-Lorenzo}},\
  }\href@noop {} {\bibfield  {journal} {\bibinfo  {journal} {Physical Review
  Letters}\ }\textbf {\bibinfo {volume} {101}},\ \bibinfo {pages} {107203}
  (\bibinfo {year} {2008})}\BibitemShut {NoStop}%
\bibitem [{\citenamefont {Newman}\ and\ \citenamefont
  {Stein}(1999)}]{newman:99}%
  \BibitemOpen
  \bibfield  {author} {\bibinfo {author} {\bibfnamefont {C.~M.}\ \bibnamefont
  {Newman}}\ and\ \bibinfo {author} {\bibfnamefont {D.~L.}\ \bibnamefont
  {Stein}},\ }\href {\doibase 10.1103/PhysRevE.60.5244} {\bibfield  {journal}
  {\bibinfo  {journal} {Phys. Rev. E}\ }\textbf {\bibinfo {volume} {60}},\
  \bibinfo {pages} {5244} (\bibinfo {year} {1999})}\BibitemShut {NoStop}%
\bibitem [{\citenamefont {{Parisi}}(1995)}]{parisi:95}%
  \BibitemOpen
  \bibfield  {author} {\bibinfo {author} {\bibfnamefont {G.}~\bibnamefont
  {{Parisi}}},\ }\href@noop {} {\bibfield  {journal} {\bibinfo  {journal}
  {eprint arXiv:cond-mat/9501045}\ } (\bibinfo {year} {1995})}\BibitemShut
  {NoStop}%
\bibitem [{\citenamefont {Roberts}(1981)}]{roberts:81}%
  \BibitemOpen
  \bibfield  {author} {\bibinfo {author} {\bibfnamefont {S.~A.}\ \bibnamefont
  {Roberts}},\ }\href {http://stacks.iop.org/0022-3719/14/i=21/a=018}
  {\bibfield  {journal} {\bibinfo  {journal} {Journal of Physics C: Solid State
  Physics}\ }\textbf {\bibinfo {volume} {14}},\ \bibinfo {pages} {3015}
  (\bibinfo {year} {1981})}\BibitemShut {NoStop}%
\bibitem [{\citenamefont {Yan}\ \emph {et~al.}(2015)\citenamefont {Yan},
  \citenamefont {Baity-Jesi}, \citenamefont {M\"uller},\ and\ \citenamefont
  {Wyart}}]{yan:15}%
  \BibitemOpen
  \bibfield  {author} {\bibinfo {author} {\bibfnamefont {L.}~\bibnamefont
  {Yan}}, \bibinfo {author} {\bibfnamefont {M.}~\bibnamefont {Baity-Jesi}},
  \bibinfo {author} {\bibfnamefont {M.}~\bibnamefont {M\"uller}}, \ and\
  \bibinfo {author} {\bibfnamefont {M.}~\bibnamefont {Wyart}},\ }\href@noop {}
  {\bibfield  {journal} {\bibinfo  {journal} {Phys. Rev. Lett.}\ }\textbf
  {\bibinfo {volume} {114}},\ \bibinfo {pages} {247208} (\bibinfo {year}
  {2015})}\BibitemShut {NoStop}%
\bibitem [{\citenamefont {Bray}\ and\ \citenamefont
  {Moore}(1981{\natexlab{b}})}]{bm:81a}%
  \BibitemOpen
  \bibfield  {author} {\bibinfo {author} {\bibfnamefont {A.~J.}\ \bibnamefont
  {Bray}}\ and\ \bibinfo {author} {\bibfnamefont {M.~A.}\ \bibnamefont
  {Moore}},\ }\href@noop {} {\bibfield  {journal} {\bibinfo  {journal} {Journal
  of Physics C: Solid State Physics}\ }\textbf {\bibinfo {volume} {14}},\
  \bibinfo {pages} {1313} (\bibinfo {year} {1981}{\natexlab{b}})}\BibitemShut
  {NoStop}%
\bibitem [{\citenamefont {M\"{u}ller}\ and\ \citenamefont
  {Wyart}(2015)}]{muller:15}%
  \BibitemOpen
  \bibfield  {author} {\bibinfo {author} {\bibfnamefont {M.}~\bibnamefont
  {M\"{u}ller}}\ and\ \bibinfo {author} {\bibfnamefont {M.}~\bibnamefont
  {Wyart}},\ }\href@noop {} {\bibfield  {journal} {\bibinfo  {journal} {Annual
  Review of Condensed Matter Physics}\ }\textbf {\bibinfo {volume} {6}},\
  \bibinfo {pages} {177} (\bibinfo {year} {2015})}\BibitemShut {NoStop}%
\bibitem [{\citenamefont {Sharma}\ and\ \citenamefont
  {Young}(2010)}]{sharma2010almeida}%
  \BibitemOpen
  \bibfield  {author} {\bibinfo {author} {\bibfnamefont {A.}~\bibnamefont
  {Sharma}}\ and\ \bibinfo {author} {\bibfnamefont {A.~P.}\ \bibnamefont
  {Young}},\ }\href@noop {} {\bibfield  {journal} {\bibinfo  {journal}
  {Physical Review E}\ }\textbf {\bibinfo {volume} {81}},\ \bibinfo {pages}
  {061115} (\bibinfo {year} {2010})}\BibitemShut {NoStop}%
\bibitem [{\citenamefont {{Lupo}}\ \emph {et~al.}(2016)\citenamefont {{Lupo}},
  \citenamefont {{Parisi}},\ and\ \citenamefont {{Ricci-Tersenghi}}}]{lupo:16}%
  \BibitemOpen
  \bibfield  {author} {\bibinfo {author} {\bibfnamefont {C.}~\bibnamefont
  {{Lupo}}}, \bibinfo {author} {\bibfnamefont {G.}~\bibnamefont {{Parisi}}}, \
  and\ \bibinfo {author} {\bibfnamefont {F.}~\bibnamefont
  {{Ricci-Tersenghi}}},\ }\href@noop {} {\bibfield  {journal} {\bibinfo
  {journal} {private communication and to be published}\ } (\bibinfo {year}
  {2016})}\BibitemShut {NoStop}%
\bibitem [{\citenamefont {Sharma}\ and\ \citenamefont
  {Young}(2011{\natexlab{a}})}]{sharma2011phase}%
  \BibitemOpen
  \bibfield  {author} {\bibinfo {author} {\bibfnamefont {A.}~\bibnamefont
  {Sharma}}\ and\ \bibinfo {author} {\bibfnamefont {A.~P.}\ \bibnamefont
  {Young}},\ }\href@noop {} {\bibfield  {journal} {\bibinfo  {journal}
  {Physical Review B}\ }\textbf {\bibinfo {volume} {83}},\ \bibinfo {pages}
  {214405} (\bibinfo {year} {2011}{\natexlab{a}})}\BibitemShut {NoStop}%
\bibitem [{\citenamefont {Sharma}\ and\ \citenamefont
  {Young}(2011{\natexlab{b}})}]{sharma2011almeida}%
  \BibitemOpen
  \bibfield  {author} {\bibinfo {author} {\bibfnamefont {A.}~\bibnamefont
  {Sharma}}\ and\ \bibinfo {author} {\bibfnamefont {A.~P.}\ \bibnamefont
  {Young}},\ }\href@noop {} {\bibfield  {journal} {\bibinfo  {journal}
  {Physical Review B}\ }\textbf {\bibinfo {volume} {84}},\ \bibinfo {pages}
  {014428} (\bibinfo {year} {2011}{\natexlab{b}})}\BibitemShut {NoStop}%
\bibitem [{\citenamefont {Boettcher}(2003)}]{boettcher:03}%
  \BibitemOpen
  \bibfield  {author} {\bibinfo {author} {\bibfnamefont {S.}~\bibnamefont
  {Boettcher}},\ }\href@noop {} {\bibfield  {journal} {\bibinfo  {journal}
  {Eur. Phys. J. B}\ }\textbf {\bibinfo {volume} {31}},\ \bibinfo {pages} {29}
  (\bibinfo {year} {2003})}\BibitemShut {NoStop}%
\bibitem [{\citenamefont {Aspelmeier}\ \emph {et~al.}(2008)\citenamefont
  {Aspelmeier}, \citenamefont {Billoire}, \citenamefont {Marinari},\ and\
  \citenamefont {Moore}}]{billoire:08}%
  \BibitemOpen
  \bibfield  {author} {\bibinfo {author} {\bibfnamefont {T.}~\bibnamefont
  {Aspelmeier}}, \bibinfo {author} {\bibfnamefont {A.}~\bibnamefont
  {Billoire}}, \bibinfo {author} {\bibfnamefont {E.}~\bibnamefont {Marinari}},
  \ and\ \bibinfo {author} {\bibfnamefont {M.~A.}\ \bibnamefont {Moore}},\
  }\href@noop {} {\bibfield  {journal} {\bibinfo  {journal} {Journal of Physics
  A: Mathematical and Theoretical}\ }\textbf {\bibinfo {volume} {41}},\
  \bibinfo {pages} {324008} (\bibinfo {year} {2008})}\BibitemShut {NoStop}%
\bibitem [{\citenamefont {Gon\c{c}alves}\ and\ \citenamefont
  {Boettcher}(2008)}]{boettcher:08}%
  \BibitemOpen
  \bibfield  {author} {\bibinfo {author} {\bibfnamefont {B.}~\bibnamefont
  {Gon\c{c}alves}}\ and\ \bibinfo {author} {\bibfnamefont {S.}~\bibnamefont
  {Boettcher}},\ }\href@noop {} {\bibfield  {journal} {\bibinfo  {journal}
  {Journal of Statistical Mechanics: Theory and Experiment}\ }\textbf {\bibinfo
  {volume} {2008}},\ \bibinfo {pages} {P01003} (\bibinfo {year}
  {2008})}\BibitemShut {NoStop}%
\bibitem [{\citenamefont {Andresen}\ \emph {et~al.}(2013)\citenamefont
  {Andresen}, \citenamefont {Zhu}, \citenamefont {Andrist}, \citenamefont
  {Katzgraber}, \citenamefont {Dobrosavljevi\ifmmode~\acute{c}\else
  \'{c}\fi{}},\ and\ \citenamefont {Zimanyi}}]{andresen:13}%
  \BibitemOpen
  \bibfield  {author} {\bibinfo {author} {\bibfnamefont {J.~C.}\ \bibnamefont
  {Andresen}}, \bibinfo {author} {\bibfnamefont {Z.}~\bibnamefont {Zhu}},
  \bibinfo {author} {\bibfnamefont {R.~S.}\ \bibnamefont {Andrist}}, \bibinfo
  {author} {\bibfnamefont {H.~G.}\ \bibnamefont {Katzgraber}}, \bibinfo
  {author} {\bibfnamefont {V.}~\bibnamefont
  {Dobrosavljevi\ifmmode~\acute{c}\else \'{c}\fi{}}}, \ and\ \bibinfo {author}
  {\bibfnamefont {G.~T.}\ \bibnamefont {Zimanyi}},\ }\href@noop {} {\bibfield
  {journal} {\bibinfo  {journal} {Phys. Rev. Lett.}\ }\textbf {\bibinfo
  {volume} {111}},\ \bibinfo {pages} {097203} (\bibinfo {year}
  {2013})}\BibitemShut {NoStop}%
\bibitem [{\citenamefont {Sharma}\ \emph {et~al.}(2014)\citenamefont {Sharma},
  \citenamefont {Andreanov},\ and\ \citenamefont
  {M{\"u}ller}}]{sharma2014avalanches}%
  \BibitemOpen
  \bibfield  {author} {\bibinfo {author} {\bibfnamefont {A.}~\bibnamefont
  {Sharma}}, \bibinfo {author} {\bibfnamefont {A.}~\bibnamefont {Andreanov}}, \
  and\ \bibinfo {author} {\bibfnamefont {M.}~\bibnamefont {M{\"u}ller}},\
  }\href@noop {} {\bibfield  {journal} {\bibinfo  {journal} {Physical Review
  E}\ }\textbf {\bibinfo {volume} {90}},\ \bibinfo {pages} {042103} (\bibinfo
  {year} {2014})}\BibitemShut {NoStop}%
\bibitem [{\citenamefont {Bray}\ and\ \citenamefont {Moore}(1979)}]{bm1979}%
  \BibitemOpen
  \bibfield  {author} {\bibinfo {author} {\bibfnamefont {A.~J.}\ \bibnamefont
  {Bray}}\ and\ \bibinfo {author} {\bibfnamefont {M.~A.}\ \bibnamefont
  {Moore}},\ }\href {http://stacks.iop.org/0022-3719/12/i=11/a=008} {\bibfield
  {journal} {\bibinfo  {journal} {Journal of Physics C: Solid State Physics}\
  }\textbf {\bibinfo {volume} {12}},\ \bibinfo {pages} {L441} (\bibinfo {year}
  {1979})}\BibitemShut {NoStop}%
\bibitem [{\citenamefont {Beton}\ and\ \citenamefont
  {Moore}(1984)}]{beton1984electron}%
  \BibitemOpen
  \bibfield  {author} {\bibinfo {author} {\bibfnamefont {P.~H.}\ \bibnamefont
  {Beton}}\ and\ \bibinfo {author} {\bibfnamefont {M.~A.}\ \bibnamefont
  {Moore}},\ }\href@noop {} {\bibfield  {journal} {\bibinfo  {journal} {Journal
  of Physics C: Solid State Physics}\ }\textbf {\bibinfo {volume} {17}},\
  \bibinfo {pages} {2157} (\bibinfo {year} {1984})}\BibitemShut {NoStop}%
\end{thebibliography}%

\end{document}